\documentclass[reprint,aps,prapplied]{revtex4-1}

\renewcommand{\Re}{\mathop{\mathrm{Re}}}

\usepackage{graphicx}
\usepackage{amsmath, amsthm, amssymb}
\usepackage{epstopdf}
\usepackage{mathtools}
\usepackage{circuitikz}
\usepackage{dcolumn}
\usepackage[tight]{subfigure}
\usepackage{amsmath}
\usepackage{verbatim}
\usepackage{inputenc}
\usepackage{color}
\usepackage{bm} 
\usepackage{natbib}
\usepackage{xspace}
\usepackage{hyperref} 
\usepackage{mathbbol}

\begin{document}

\title{A superconducting nonlinear thermoelectric heat engine}
\author{G. Marchegiani}
\email{giampiero.marchegiani@nano.cnr.it} 

\affiliation{NEST Istituto Nanoscienze-CNR and Scuola Normale Superiore, I-56127 Pisa, Italy}

\author{A. Braggio}
\email{alessandro.braggio@nano.cnr.it} 
\affiliation{NEST Istituto Nanoscienze-CNR and Scuola Normale Superiore, I-56127 Pisa, Italy}

\author{F. Giazotto}
\email{francesco.giazotto@sns.it} 
\affiliation{NEST Istituto Nanoscienze-CNR and Scuola Normale Superiore, I-56127 Pisa, Italy}

\date{\today}

\begin{abstract}
In a previous work, we predicted that a thermally biased tunnel junction between two different superconductors can display a thermoelectric effect of nonlinear nature in the temperature gradient, under proper conditions. In this work we give a more extended discussion, and we focus on the two main features of the nonlinear contributions: i) the \textit{linear-in-bias} thermoelectricity, that can be associated to a spontaneous breaking of electron-hole symmetry, ii) the strong contribution at the matching peak singularity, which is typically associated to the maximum output power and efficiency. We discuss the nonlinear origin of the thermoelectricity and its relationship with the non-linear cooling mechanism in superconducting junctions previously discussed in the literature. Finally, we design and characterize the performance of the system as a heat engine, for a realistic design and experimental parameter values. We discuss possible non-idealities demonstrating that the system is amenable to current experimental realization.
\end{abstract}

\maketitle
\section{Introduction}
The degree of control of nano-fabrication techniques reached over the last few decades has stimulated the investigation of thermal transport at the micro-nanoscale~\cite{Benenti2017,Goodson2006,GiazottoRMP,Muhonen2012,DiVentraRMP,KosloffEntropy,CahillJAP2003,BergfieldMolecular,WangEPJB,Pop2010}. On the theoretical side, the interest range from the investigation of exotic nonequilibrium phenomena, and quantum effects on the thermodynamical laws~\cite{Benenti2017}. From the experimental side, there has been a strong effort in the development of on-chip coolers~\cite{Chowdhury2009,SciRepBradley2017,Ziabari2016,Shakouri2006,GiazottoRMP,Muhonen2012,PrancePRL}, and the possibility of making use of unwanted waste energy~\cite{Bell2008,Sothmann_2014,QDHarvesterPRL,Thierschmann2015,Roche2015,HartmannPRL2015}. In this direction, thermoelectric elements may play a crucial role, thanks to the direct heat-to-current conversion~\cite{Goldsmid2016}. There is currently an extensive literature concerning the theoretical modeling of thermoelectric devices~\cite{Benenti2017}, with first investigation in the nonlinear regime~\cite{SanchezNonlinearReview}, and few experimental implementations~\cite{Reddy2007,Brantut2013,Roche2015,Thierschmann2015,Josefsson2018}. In this context, superconducting junctions plays an important role, due to their consolidated fabrication process and their massive use in quantum technologies~\cite{Kurizki3866} and qubits~\cite{PekolaReview2015,Wendin2017,Krantz2019}. They have been successfully used for cooling purposes~\cite{GiazottoRMP,Muhonen2012} and for the coherent control of heat currents~\cite{FornieriReview,Hwang2020}. Very recently, they have been also used, in combination with ferromagnetic elements, to generate strong~\cite{BelzigTEPRL,Ozaeta2014,Kolenda2017,Beckmann2016} or nonlocal thermoelectric effects~\cite{BraggioNonlocalTE,GlazmanPRB99,LesovikPRB99,Blasi}. They can be used as local thermometers~\cite{GiazottoThermometerNFIS},
 for wireless delivery of power~\cite{MarchegianiEngine}, for autonomous refrigeration~\cite{MarchegianiCooler}, and for sensitive radiation detection~\cite{GiazottoTEDetector}. This technology seems really promising but it is also challenging from the experimental side~\cite{Strambini2017,DeSimoni2018}, due to the excellent quality requested in the ferromagnetic-superconducting contacts. In a previous work, we have demonstrated that, even in the absence of spin-dependent mechanism, superconducting junctions can displays strong thermoelectric effects in the nonlinear regime~\cite{MarchegianiNLTE}. This is a striking result, since the nearly perfect electron-hole symmetry of superconductors makes linear thermoelectric effects negligible. The purpose of this work is to give a more extended discussion of thermoelectricity in superconducting tunnel junctions, and its main features. Moreover, we present a design study for a possible proof of principle demonstration of this nonlinear thermoelectricity and the actual implementation of a heat engine based on the superconducting technology using realistic parameters.
\section{Two terminal thermoelectricity} 
\label{sec:model}
\begin{figure}
    \includegraphics[width=0.5\textwidth]{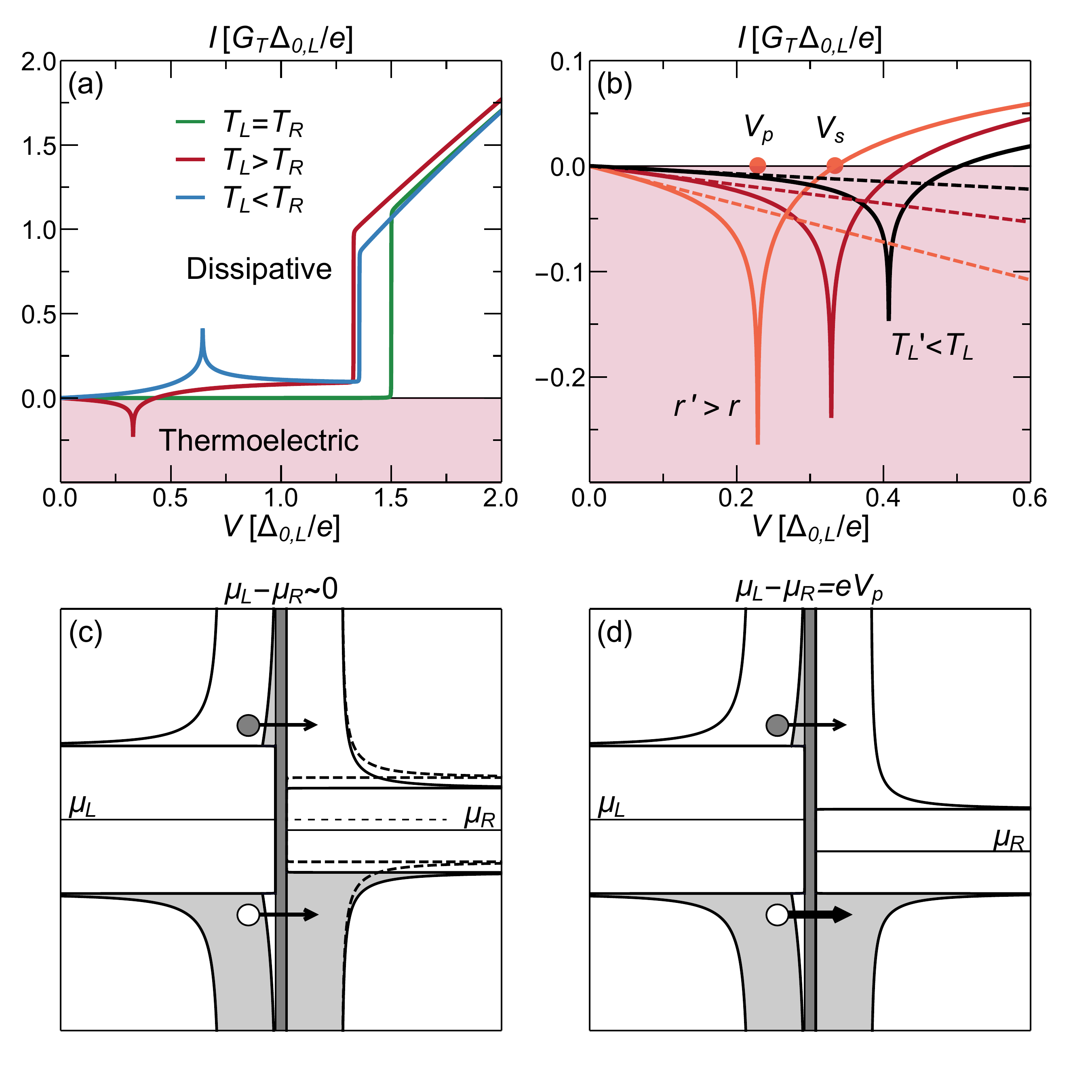}
    \caption{(a) Quasiparticle current-voltage characteristic of a tunnel junction between two different superconductors ($L,R$), with $r=\Delta_{ 0,R}/\Delta_{0,L}<1$, for different temperature biases. Parameters are $r=0.5$, $\Gamma_{ \alpha}/\Delta_{ 0,\alpha}=10^{-4}$ and $T_{ L}=T_{R}=0.1 T_{c,L}$ (green), $T_{L}=0.7 T_{ c,L},T_{R}=0.1 T_{c,L}$ (red), $T_{L}=0.1 T_{ c,L},T_{R}=0.4 T_{c,L}$ (blue). A negative current $I(V>0)<0$ characterizes a thermoelectric behavior (pink area). (b) Magnification of the $I(V)$ curve at small values of the bias. The red curve is the same of panel a). The black curve is obtained from the red by decreasing the hot temperature to $T_{L}=0.6 T_{c,L}$ and the orange by setting $r=0.6$. In the orange curve, the values of the matching peak bias $V_{p}$ and the stopping voltage $V_{s}$ are explicitly drawn with orange points. c)-d) Description of the thermoelectricity in the energy band diagram. (c) The \textit{linear-in-bias} thermoelectricity arises since the hole current (hollow circles) overcomes the particle current (filled circles) due to the local monotonically decreasing density of states of the cold electrode ($R$). (d) Enhancement of the thermoelectric current due to the matching of the singularity peaks of the superconducting density of states (\textit{nonlinear-in-bias} contribution).}
    \label{Fig1}
\end{figure}

We consider a tunnel junction between two Bardeen-Cooper-Schrieffer (BCS~\cite{Tinkham2004}) superconductors ($L,R$) and assume each electrode in internal thermal equilibrium, namely the quasiparticle distributions read $f_\alpha(E-\mu_\alpha)=\{1+\exp[(E-\mu_\alpha)/(k_{ B}T_\alpha)]\}^{-1}$, where $k_{ B}$ is the Boltzmann constant and $T_\alpha$, $\mu_\alpha$ (with $\alpha=L, R$) are the temperatures and the chemical potentials of the quasiparticle systems, respectively. We focus on the quasiparticle transport across the junction and we completely disregard the contributions due to the Josephson effect~\cite{Tinkham2004,barone1982physics}. This latter condition can be achieved experimentally in different ways. For instance, the Josephson current is suppressed by applying a strong in-plane magnetic field or by applying a small out-of-plane magnetic field in a direct-current superconducting quantum interference device (dc-SQUID)~\cite{Tinkham2004,barone1982physics,GiazottoRMP}. Another possibility involves the use of strongly oxidized tunnel barriers, where the Josephson coupling energy $E_J$ is destroyed by thermal fluctuations since $E_J\ll k_BT$ (here $T=(T_L+T_R)/2$)~\cite{barone1982physics}.

Hence, the transport is completely associated to quasiparticles, and the charge and the heat current flowing out of the $\alpha$-electrode (with $\bar\alpha={ R}$ when $\alpha={ L}$ and \textit{vice versa}) read~\cite{GiazottoRMP}
\begin{equation}
\begin{pmatrix}
I_{\alpha}   \\   
\dot Q_{\alpha}
\end{pmatrix}
=\frac{G_{ T}}{e^2}\int_{-\infty}^{+\infty}dE 
\begin{pmatrix}
-e \\   
E_\alpha
\end{pmatrix} N_{\alpha}(E_{\alpha})N_{\bar\alpha}(E_{\bar\alpha})F_\alpha(E_\alpha)
\label{eq:IVandQ}
\end{equation}
where $-e$ is the electron charge, $N_\alpha(E)=|\Re[(E+i\Gamma_\alpha)/\sqrt{(E+i\Gamma_\alpha)^2-\Delta^2_\alpha}]|$ is the smeared (by the Dynes parameter $\Gamma_\alpha\ll\Delta_\alpha$~\cite{Dynes1978,Dynes1984})  quasiparticle density of states (DOS), $F_\alpha(E_\alpha)=f_\alpha(E_\alpha)-f_{\bar\alpha}(E_{\bar\alpha})$, $E_\alpha=E-\mu_\alpha$, and $G_{ T}$ is the normal-state conductance of the junction. In the BCS model, the energy gap $\Delta_\alpha(T_{\alpha})$ is a monotonically decreasing function, and it is zero for temperatures larger than the critical temperature $T_{ c,i}=\Delta_{0,i}/(1.764 k_{ B})$~\cite{Tinkham2004}, where $\Delta_{0,i}$ is the zero-temperature value (with $i=$L,R). For our purposes, we define the ratio between the two zero-temperature values as $r=\Delta_{ 0,R}/\Delta_{ 0,L}$, which is associated to the asymmetry of the two superconductors. With no loss of generality, we consider $r\leq 1$ in this work.
Since $N_{\alpha}(E_\alpha)=N_{\alpha}(-E_\alpha)$, the system displays electron-hole (EH) symmetry and it results $I(V,T_{L},T_{R})=-I(-V,T_{L},T_{R})$ where $I=I_{L}$ and $V=(\mu_{L}-\mu_{R})/(-e)$ is the voltage bias across the junction. 
In the linear response regime, i.e., for a small voltage bias and a small temperature bias, thermoelectric effects vanish due to this symmetry~\cite{Benenti2017,MarchegianiNLTE}. The situation change in the presence of a nonlinear temperature bias, as we firstly discussed in Ref.~\cite{MarchegianiNLTE}. In particular, we demonstrated that an asymmetric junction between two superconductors (S'IS junction), i.e., for $r\neq 1$, can display a finite thermoelectric power $\dot W=-IV>0$, under proper conditions. This nonlinear thermoelectricity corresponds to the esistence of an absolute negative conductance $G(V,T_{ L},T_{ R})=I(V,T_{ L},T_{ R})/V<0$, which can occur when a temperature difference $T_{ L}\neq T_{ R}$ is applied to the junction. Note that the absolute negative conductance in tunnel junctions between two superconductors has been already predicted~\cite{Aronov1975} and demonstrated in non-equilibrium experiments with particle injection~\cite{Gershenzon1986,Gershenzon1988,Gijsbertsen1996} and microwave irradiation~\cite{NagelPRL}. However, the thermoelectric effect here discussed has not been investigated, yet.

Figure~\ref{Fig1}a displays the shape of the current-voltage characteristic for $r=0.5$ and different values of the temperatures of the two electrodes. Thanks to the EH symmetry, we can discuss only the case $V>0$. A positive current ($I>0$) denotes a dissipative behavior ($\dot W<0$), whereas a negative current ($I<0$) corresponds to a thermoelectric generation (pink area). For equal temperatures $T_{ L}=T_{ R}=T$, the junction is always dissipative since a positive value of $\dot W$ would imply a negative entropy rate production $\dot S<0$, and a violation of the second law of thermodynamics~\cite{Benenti2017,MarchegianiNLTE}. In particular, for $k_{ B}T\ll \Delta_{ 0,L},\Delta_{ 0,R}$ (green curve), the transport is strongly suppressed at subgap voltages, i.e., $I\sim \Gamma_{ L}\Gamma_{ R}G_{ T}V/(\Delta_{ 0,L}\Delta_{ 0,R})$ for $eV<\Delta_{ 0,L}+\Delta_{ 0,R}$, and it is almost linear at larger values $eV\gg \Delta_{ 0,L}+\Delta_{ 0,R}$, where it asymptotically reads $I\sim G_{ T}V$. In the presence of a strong temperature difference between the electrodes, the evolution is more complex. In particular, $I(V)$ is non-monotonic and shows a characteristic peak at $V_{ p}=\pm|\Delta_{ L}(T_{ L})-\Delta_{ R}(T_{ R})|/e$, due to the matching of the BCS singularities. While for $T_{ L}<T_{ R}$ the junction is dissipative (blue curve), when $T_{ L}>T_{ R}$ (red curve), i.e., when the larger gap superconductor is heated up, the curve may display a region of absolute negative conductance and thermoelectricity provided that $\Delta_{ L}(T_{ L})>\Delta_{ R}(T_{ R})$~\cite{MarchegianiNLTE}. 

The typical subgap voltage evolution in the presence of thermoelectricity is displayed in Fig.~\ref{Fig1}b. 
The red curve is a magnification of the $T_{ L}>T_{ R}$ curve of Fig.~\ref{Fig1}a. The other curves differs from the red due to a single parameter modification. 
In particular, in the black curve the temperature of the hot electrode is slightly decreased ($T_{ L}\rightarrow T_{ L}'<T_{ L}$) while in the orange curve the symmetry parameter is slightly increased $r\rightarrow r'>r$. We firstly note that the curves display an almost linear behaviour with a negative slope at a small voltage bias, i.e., $I(V,T_{ L},T_{ R})\sim g_0(T_{ L},T_{ R})V$, where the zero-bias differential conductance $g_0(T_{ L},T_{ R})=\partial I(V,T_{ L},T_{ R})/\partial V|_{V=0}$ is negative (see dashed lines in Fig.~\ref{Fig1}b). 
This negative slope shows that the system presents a \textit{linear-in-bias} thermoelectric contribution in the presence of a \textit{nonlinear} temperature difference. Since the junction recovers a dissipative behaviour characterized by a positive conductance at sufficiently high voltage bias (in particular $I(V)/V\sim G_{ T}$ for $V\gg(\Delta_{ L}+\Delta_{ R})/e$), this \textit{linear-in-bias} contribution implies the existence of, at least, a point ($V_{ s}\neq 0$), where the current is zero, i.e., $I(V_{ s})=0$ (see Fig.~\ref{Fig1}b, showing only the positive bias side). This finite value $V_{s}$ (note that a similar behavior occurs at $-V_{s}$, due to the EH symmetry) is also called Seebeck voltage, and represents the value where the intrinsic thermoelectricity of the junction is no longer able to counteract the electric transport due to the voltage bias. Note that the two values of the Seebeck voltage are both possible for the \emph{same} temperature gradient. In addition, this negative differential conductance ($g_0<0$) implies an electrical instability at the zero current state with $V=0$~\cite{MarchegianiNLTE}. Namely, any spurious fluctuation of the voltage around $V=0$ drives the junction in the zero-current state with a finite thermoelectric voltage (either $\pm V_{ s}$). In other words, the EH symmetry is spontaneously broken due to the presence of the nonlinear temperature difference. 

In Fig.~\ref{Fig1}b the matching peak value $V_{ p}$ appears at intermediate values of the applied bias, i.e., for $V\lesssim V_{ s}$ and represents the condition where the absolute value of the thermoelectric current and the thermoelectric power reach their maximum value, i.e., $\dot W_{\rm max}=\max_{V}(-IV)\sim -I(V_{ p})V_{ p}$. This condition represents the main \textit{nonlinear-in-bias} contribution on thermoelectricity. In Fig.~\ref{Fig1}b, the position of the matching peak changes by modifying either $T_{L}$ or $r$. In particular, by decreasing $T_{L}$ (black curve), $V_{p}$ shifts towards higher voltages (since at the same time $\Delta_\alpha(T_{L})$ increases). Intriguingly, this shows the peculiar nonlinear nature of the thermoelectricity, where the absolute value of the Seebeck voltage increases by slightly decreasing the temperature gradient. Similarly, upon increasing $r$ (orange curve), $V_{p}$ shifts towards lower voltages. Note that the \textit{linear-in-bias} contribution, which is characterized by $g_0$ (slope of the dashed line), is modified as well. More precisely, $|g_0|$ increases when $V_{p}$ decreases (for $r'>r$, orange curve) and \emph{vice versa} (for $T_{L}'<T_{ L}$, black curve).

In summary, the nonlinear thermoelectricity in the S'IS junction is characterized by two main contributions, namely the \textit{linear-in-bias} and the \textit{nonlinear-in-bias}. The origin of the thermoelectric effect for $T_{ L}>T_{ R}$ can be intuitively understood in the semiconductor model, as displayed in Fig.~\ref{Fig1}c-d. For simplicity, we discuss the behaviour of the particle current $I/(-e)$ and we consider the case $T_{ R}\rightarrow 0^+$. The current from L to R is the difference between the particle current above the chemical potential $\mu_{ L}$ (filled circles) and the hole current below the chemical potential $\mu_{ R}$ (empty circles). First, we focus on the \textit{linear-in-bias} contribution. For $V=0$, the two chemical potentials are aligned $\mu_{ L}=\mu_{ R}$ (see dashed lines in Fig.~\ref{Fig1}c) and the particle and the hole contributions cancel out due to the EH symmetry.
In the presence of a voltage bias, the chemical potential are shifted with respect to each other. In particular, let's focus on $\mu_{ L}>\mu_{ R}$, where the particle current naturally flows from L to R in the standard (dissipative) regime. Note that, due to the monotonically decreasing DOS of the right electrode above gap, i.e., for $E>\mu_{ R}+\Delta_{ R}$ (and hence monotonically increasing for $E<\mu_{ R}-\Delta_{ R}$ due to EH symmetry), the particle current contribution is decreased due to the shift, whereas the hole contribution is increased. As a consequence, the system displays a negative particle current. The unbalance is maximized when $\mu_{ L}-\mu_{ R}=\Delta_{ L}-\Delta_{ R}$, due to the matching of the BCS singularities (see Fig.~\ref{Fig1}d). This scheme also explains why the thermoelectric effect is absent for $T_{ L}<T_{ R}$. In this case, the right electrode is the hotter one, and therefore the arrows in Fig.~\ref{Fig1}c)-d) must be drawn necessarily in the opposite direction. However, it is still true that the hole contribution is larger than the particle contribution for subgap biases. Hence, the particle current flows in the direction of the chemical potential gradient (from L to R), the system becomes dissipative and no thermoelectricity is possible. In summary, the semiconductor model clearly shows that the nonlinear thermoelectricity is obtained in the presence of two conditions: i) the larger gap electrode should be heated up, ii) the colder electrode must have a local monotonically decreasing DOS. In a S'IS junction, these two conditions clearly show that, for $r\leq 1$ (as assumed in this work), themoelectricity arises only for $T_{ L}>T_{ R}$ provided that the hot electrode has the largest gap, namely $\Delta_{ L}(T_{ L})>\Delta_{ R}(T_{ R})$. In the next section we give a more quantitative discussion of the nonlinear thermoelectricity, and we discuss the role of the various parameters.

\section{Nonlinear Thermoelectricity}
\begin{figure}
    \includegraphics[width=0.5\textwidth]{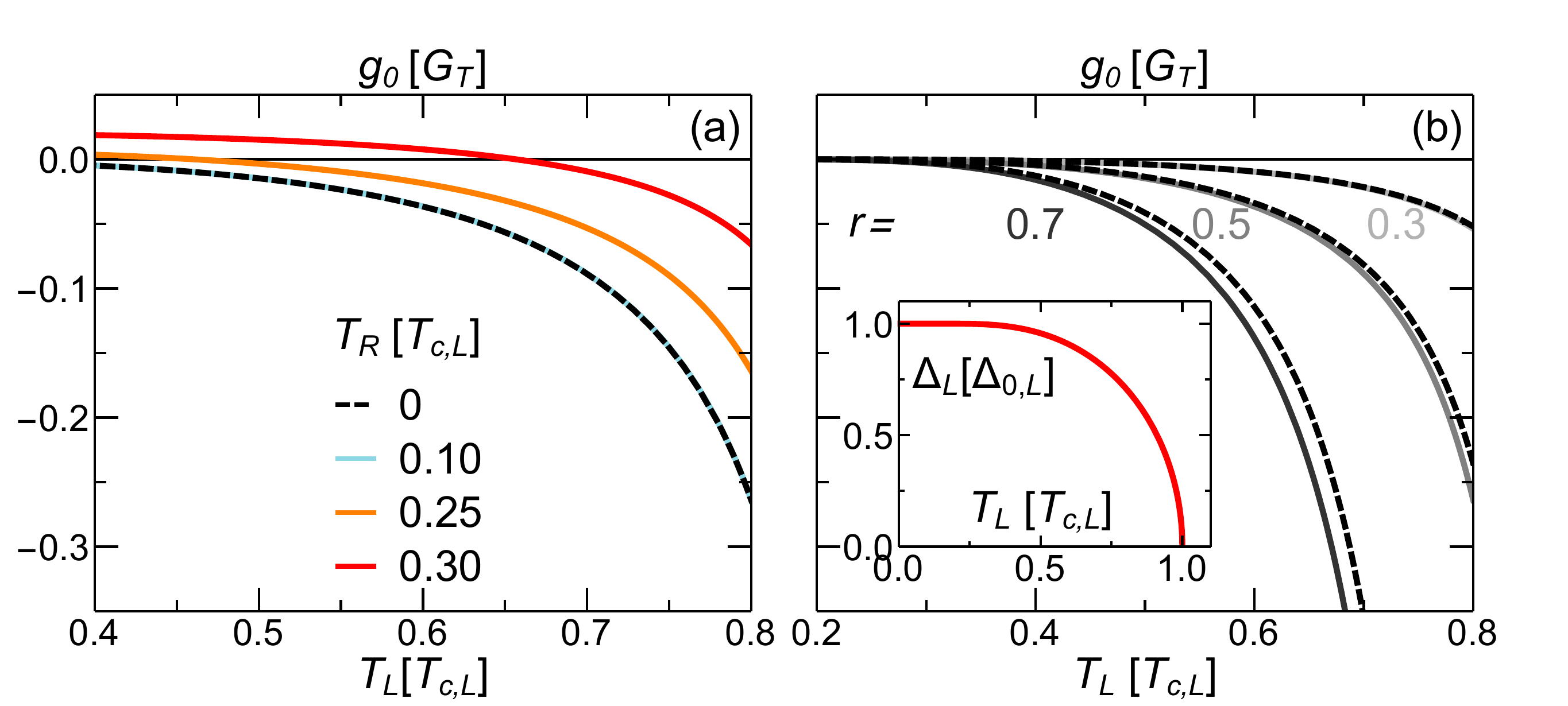}
    \caption{Zero-bias differential conductance of the junction as a function of the temperature of the hot electrode $T_{L}$. (a) Zero-bias differential conductance obtained through numerical differentiation of the charge current for $r=0.5$ and different values of the cold electrode temperature $T_{R}$ (solid). The dashed curve gives the approximate expression for $T_{ R}\rightarrow 0$ of Eq.~\ref{eq:G0}. (b) Zero-temperature limit for the cold electrode $T_{R}\rightarrow 0$ and different values of $r$ (grayscale). Solid lines are expressed by Eq.~\ref{eq:G0} and the dashed lines gives the low-$r$ approximation of Eq.~\ref{eq:G0int}. Inset: temperature dependence of the superconducting gap in the BCS weak coupling limit. In the numerics, $\Gamma_{ \alpha}/\Delta_{ 0,\alpha}=10^{-4}$.}
    \label{Fig2}
\end{figure}

In the previous section, we have qualitatively discussed the origin of the thermoelectric effect, which relies on the competition between the particle and the hole current. Here, we give a quantitative discussion and we neglect any effect associated with the Dynes parameter, for simplicity. At subgap voltages $eV<\Delta_{ L}+\Delta_{ R}$, the latter is described by the formula (with $E_\pm=E\pm eV$)
\begin{equation}
I\sim\frac{G_{ T}}{e}\int_{\Delta_{ L}(T_{ L})}^{\infty}dE N_{ L}(E)f_{ L}(E)[N_{ R}(E_+)-N_{ R}(E_-)],
\label{eq:Iapprox}
\end{equation}
which yields a good approximation in the limit $k_{ B}T_{ R}\ll \Delta_{ R}(T_{ R}),\Delta_{ L}(T_{ L})$~\cite{MarchegianiNLTE}, neglecting corrections of order $\sim\exp(-\Delta_{ 0,R}/k_{ B}T_{ R})$ and becomes exact in the limit $T_{ R}\rightarrow 0$. This expression is derived from the first of Eq.~\ref{eq:IVandQ} through a series of transformations based on the EH symmetry of the density of states $N_{\alpha}$ and on the identity $f_{ L}(E)=1-f_{ L}(-E)$~\cite{MarchegianiNLTE}.
From Eq.~\ref{eq:Iapprox}, one can obtain the two conditions for the nonlinear thermoelectricity presented in the previous section~\cite{MarchegianiNLTE}. Furthermore, one can compute the behavior at $V\sim 0$ and hence $g_0$, which characterizes the $\textit{linear-in-bias}$ thermoelectricity. In particular, in the presence of a nonlinear temperature gradient, i.e., for a finite value of $T_{ L}$, the zero-bias differential conductance is negative and reads 
\begin{equation}
g_{0}=-2G_{ T}\Delta_{ 0,R}^2\int_{\Delta_{ L}(T_{ L})}^{\infty}dE \frac{N_{ L}(E)f_{ L}(E)}{(E^2-\Delta_{ 0,R}^2)^{3/2}},
\label{eq:G0}
\end{equation}
valid for $T_{ R}=0$ and $\Gamma\rightarrow 0$, provided that $\Delta_{ L}(T_{ L})>\Delta_{ 0,R}$. 
The goodness of the low-$T_{ R}$ expression of Eq.~\ref{eq:G0} is investigated in Fig.~\ref{Fig2}a, where the temperature evolution of $g_0$ (computed through numerical differentiation of the charge current in Eq.~\ref{eq:IVandQ}) is displayed for $r=0.5$ and different values of $T_{ R}$ (solid curves). As discussed above, the approximate expression of Eq.~\ref{eq:G0} (dashed curve), which does not depend explicitly on $T_{ R}$, gives a good approximation for $T_{ R}\lesssim 0.2 T_{ c,L}$, but it is inaccurate at large values of $T_{ R}$, where the approximations which lead to Eqs.~\ref{eq:Iapprox},\ref{eq:G0} don't apply anymore. Note that, for $T_{R}\geq 0.2 T_{ c,L}$, the zero-bias conductance is positive if the temperature of the hot electrode $T_{ L}$ is smaller than a threshold value. This is related to the unavoidable nonlinear nature of the thermoelectric effect. 

In the limit $r\ll1$, the zero-bias differential conductance of Eq.~\ref{eq:G0} is well described by the interpolation formula
\begin{equation}
g_0^{\rm approx}(T_{ L})\sim-0.89\frac{r^2[1-\Delta_{ L}(T_{ L})/\Delta_{ 0,L}]}{[\Delta_{ L}(T_{ L})/\Delta_{ 0,L}]^2-r^2}G_{ T}.
\label{eq:G0int}
\end{equation}

The degree of validity of this expression is displayed in Fig.~\ref{Fig2}b, where the temperature evolution of $g_0$ of Eq.~\ref{eq:G0} (solid curves) is compared with the simplified expression of Eq.~\ref{eq:G0int} for some values of $r$. Note that the approximation becomes exact for $T_{ R},r\rightarrow 0$ but still well represents the overall behavior of the function and it is reasonably accurate also for relatively large values of $r$, i.e., $r=0.7$. Moreover, it explains the qualitative behaviour of the curves in Fig~\ref{Fig2}b. In particular, for a given $r$, $g_0$ is negative, monotonically decreasing with $T_{ L}$ and it quite small if $T_{ L}\ll T_{ c,L}$. This behavior is related to the term $1-\Delta_{ L}(T_{ L})/\Delta_{ 0,L}$ in the numerator of Fig~\ref{Fig2}b, where $\Delta_{ L}(T_{ L})$ is displayed in the inset of Fig~\ref{Fig2}b. Note that $g_0$ diverges for the temperature value where $\Delta_{ L}(T_{ L})=\Delta_{ 0,R}$, which annihilates the denominator of Eq.~\ref{eq:G0int}.

\begin{figure}
    \includegraphics[width=0.5\textwidth]{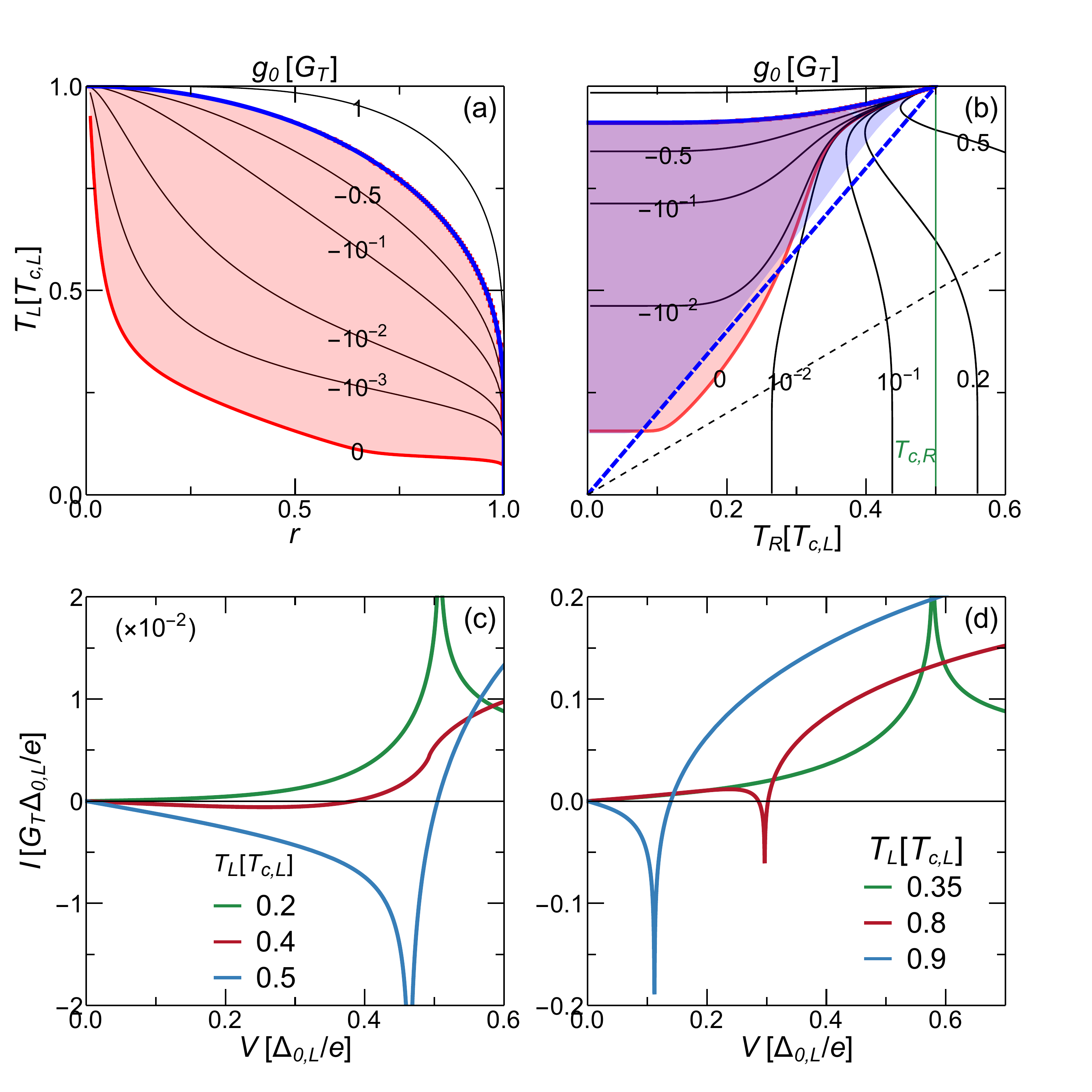}
    \caption{(a) Contour plot of $g_0$ vs $T_{ L}$ and $r$ for $T_{ R}=0.001 T_{ c,L}$. The \textit{linear-in-bias} thermoelectric contribution is represented by the red area.
    (b) Contour plot of $g_0$ vs $T_{ L}$ and $T_{ R}$ for $T_{ R}=0.001 T_{ c,L}$. The red area and the blue area denote the \textit{linear-in-bias} thermoelectric region and the \textit{nonlinear-in-bias} thermoelectric region, respectively. The dashed lines give the zero temperature difference contour $T_{ L}=T_{ R}$ (black) and the contour  $T_{ L}=T_{ R}/r$ (blue). (c)-(d) On-set of the thermoelectricity by raising the temperature of the left electrode for $r=0.5$ and (c) $T_{ R}=0.2T_{ c,L}$ (first \textit{linear-in-bias} then \textit{nonlinear-in-bias}) or (d) $T_{ R}=0.35T_{ c,L}$ (first \textit{nonlinear-in-bias} then \textit{linear-in-bias}).}
    \label{Fig3}
\end{figure}
We wish now to give a more complete discussion on the conditions where the \textit{linear-in-bias} thermoelectricity appears for $T_{ R}\rightarrow 0$. In this respect, Fig.~\ref{Fig3}a displays the contour plot of $g_0$ as a function of $T_{ L}$ and $r$ for a very low temperature of the right lead $T_{ R}=0.001T_{c,L} $. As discussed above, the thermoelectric region is characterized by $g_0<0$ (red area). For a given value of $r\leq 1$, thermoelectricity arises only if the temperature of the hot electrode $T_{L}$ is larger than a threshold value, which is represented by the $g_0=0$ contour in Fig.~\ref{Fig3}a (red curve) and smaller than an upper threshold value, where $\Delta_{ L}(T_{ L})=\Delta_{ 0,R}$ (blue curve). In the latter, the differential conductance switches very rapidly from large negative values to large positive values, due to the matching of the BCS singularities. On the other hand, the lower threshold value is due to the finite subgap conductance and cannot be captured by the expression of Eq.~\ref{eq:G0}, which is derived for $\Gamma\rightarrow 0^+$.
Additional considerations can be made in the complementary description given in Fig.~\ref{Fig3}b, where the contour plot of $g_0$ is displayed as a function of the temperature of the two electrodes for $r=0.5$. In the figure we also compare the \textit{linear-in-bias} and the \textit{nonlinear-in-bias} contributions to the nonlinear thermoelectricity.
In particular, the red area denotes the region of \textit{linear-in-bias} thermoelectricity $g_0<0$, whereas the blue region gives the \textit{nonlinear-in-bias} thermoelectricity $G_{p}=G(V_p,T_L,T_R)=I(V_p,T_L,T_R)/V_p<0$. On the dashed line the temperature difference is zero ($T_{L}=T_{R}$). 
Several features can be easily captured from the plot. First, there is no thermoelectric effect (white regions): i) for $T_{R}>T_{L}$, i.e., heating the larger gap superconductor is a necessary condition for thermoelectricity, ii) for $T_{L}\lesssim 0.2 T_{ c,L}$, due to the subgap contribution to the current related to the finite Dynes parameter, iii) for values of $T_{L}$, where $\Delta_{L}(T_{L})<\Delta_{R}(T_{R})$ (above the  blue solid curve), iv) for $T_{R}>T_{c,R}=0.5 T_{ c,L}$ (irrespectively of $T_{L}$), since the right electrode is in the normal state. The last point is associated to the fact that the thermoelectric effect cannot be observed in a hybrid normal-superconducting tunnel junction. Indeed, in the thermoelectric effect we discuss it is crucial the monotonically decreasing DOS of the right electrode above gap, which is guaranteed by the superconducting state, as previously discussed in Ref.~\cite{MarchegianiNLTE}. In a normal metal, the DOS is energy independent on the relevant energy scale, i.e., $\Delta_{0,L}$, which is much smaller than the Fermi energy. This can be intuitively understood also in the representation of Fig.~\ref{Fig1}c-d by replacing the BCS DOS in the right electrode with a flat distribution. In summary, with respect to the thermoelectric effect discussed in this work, there is nothing special about the superconducting state of the cold electrode rather then the locally monotonically decreasing DOS. In other words, any system which presents a monotonically decreasing DOS in the cold lead, a gapped DOS in the hot lead and has an EH symmetry around the chemical potential would support a nonlinear thermoelectricity similar to the one discussed here.

Secondly, a nonlinear temperature gradient is requested for thermoelectricity. In fact, in Fig.~\ref{Fig3}b the thermoelectricity is typically present \emph{only} away from the equal temperature condition $T_{ L}=T_{ R}$ (black dashed line). The numerical calculations show that the critical value of $T_L$ for the onset of the \textit{nonlinear-in-bias} thermoelectricity is roughly given by $T_R/r$ (see blue dashed line in Fig.~\ref{Fig3}). Finally, the plots show that it is possible to have \textit{linear-in-bias} thermoelectricity even in the absence of \textit{nonlinear-in-bias} thermoelectricity (red curve in Fig.\ref{Fig3}c), i.e., when the junction at the matching peak value is still dissipative, and \textit{vice-versa} (red curve in Fig.\ref{Fig3}d). The on-set of the thermoelectric effect upon increasing the temperature difference in these two particular cases (obtained for two different values of $T_{ R}$) are shown in Fig.~\ref{Fig3}c and Fig.~\ref{Fig3}d, respectively. In Fig.~\ref{Fig3}c, $T_{ R}=0.2T_{ c,L}$ and the \textit{linear-in-bias} thermoelectricity arises  when $T_{ L}\gtrsim0.3 T_{ c,L}$ (see Fig.~\ref{Fig3}b). Note that, in the transition which leads to the on-set of nonlinear thermoelectricity (for $T_{ L}\gtrsim0.43 T_{ c,L}$, see Fig.~\ref{Fig3}b), the matching peak changes the direction of the cuspid by passing through a flex. In Fig.~\ref{Fig3}d, $T_{ R}=0.35T_{ c,L}$, the nonlinear-in-bias contribution appears even in the absence of linear-in-bias thermoelectricity. In this case, there is no electrical instability of the zero current state with $V=0$ since $g_0>0$. As a consequence, a finite bias is requested to drive the system in the thermoelectric state where the instability would bring the system to the zero-current solution with $V_{s}\neq 0$.

\begin{figure}
    \includegraphics[width=0.5\textwidth]{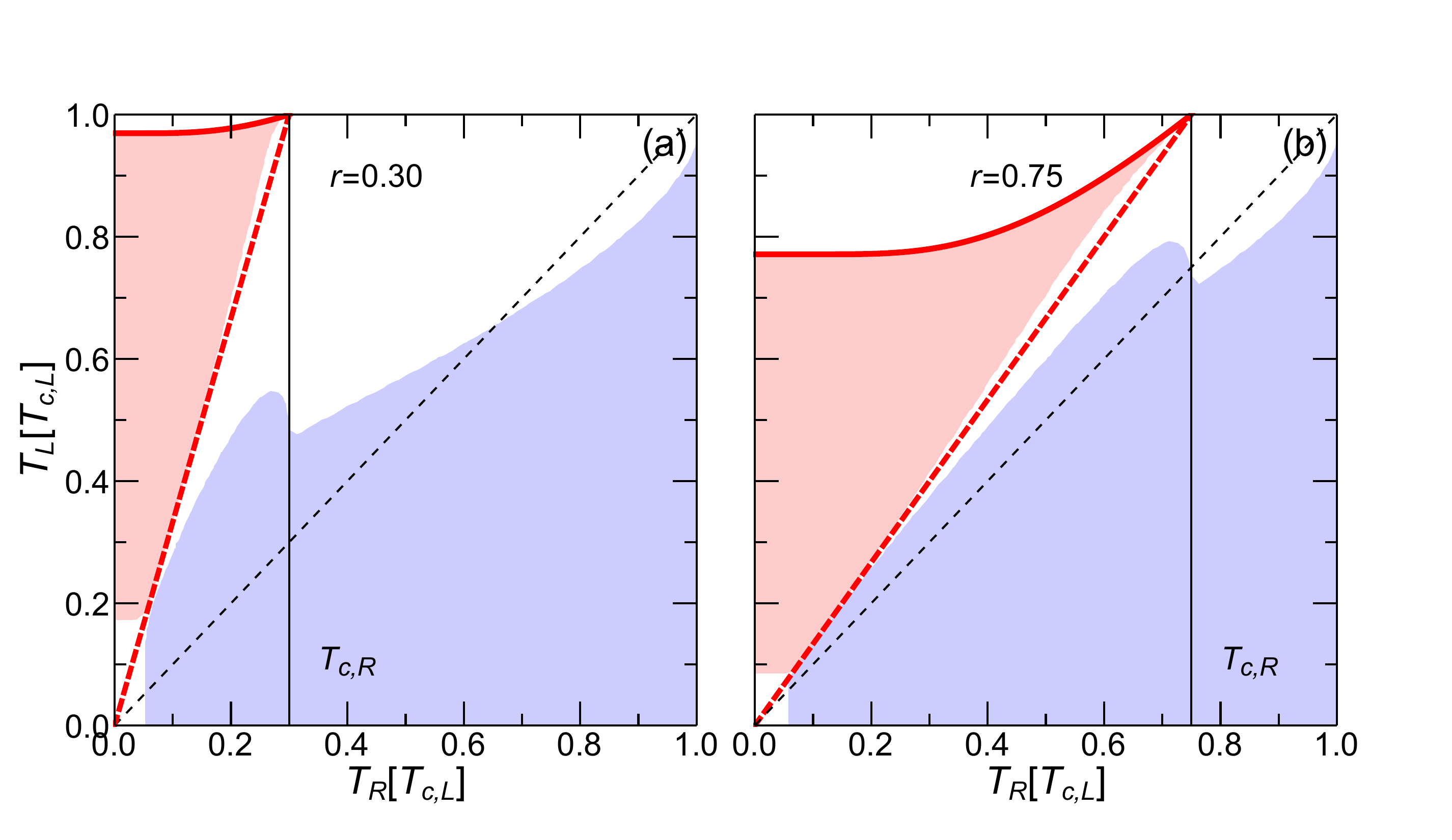}
    \caption{Competition between the nonlinear thermoelectric effect (red area) and evaporative cooling (blue area) in a voltage bias superconducting tunnel junction (with $V=V_{ p}$) for (a) $r=0.3$ and (b) $r=0.75$. The dashed lines give the zero temperature difference contour $T_{ L}=T_{ R}$ (black) and the contour  $T_{ L}=T_{ R}/r$ (red).}
    \label{Fig4}
\end{figure}

\subsection{Thermoelectricy and cooling}
To conclude this section, we discuss the relationship between the \textit{nonlinear-in-bias} thermoelectric effect in our structure and the evaporative cooling in superconducting tunnel junctions. In fact, it is well known~\cite{GiazottoRMP,Muhonen2012} that it is possible to achieve cooling of the electronic temperature of a normal conductor in a tunnel junction between a normal metal and a superconductor (NIS junction). In particular, for a NIS junction this mechanism is known as NIS cooling, and it is based on the energy filtering provided by the superconducting gap. A similar mechanism is also discussed for S'IS junctions, which we are discussing,  with $r\neq 1$, where one can achieve refrigeration of the lower gap superconductor~\cite{GiazottoRMP,Muhonen2012}. Namely, in our notation, it is possible to have cooling power $\dot Q_{R}>0$ for $T_{ L}\geq T_{ R}$, provided $r<1$. Hence, the thermoelectric effect discussed in this work and the evaporative cooling share some similarities:
i) they require the condition $\Delta_{L}>\Delta_{ R}$,
ii) they require a finite voltage bias $V$,
iii) the maximum performance in terms of cooling power/thermoelectric power are achieved for $V=V_{ p}$.
Indeed, these two effects are somewhat complementary since they cannot coexist due to the thermodynamical laws. In fact, the cooling power reads $\dot Q_{ R}=\dot{W}-\dot Q_{L}$ due to the energy conservation. In a thermoelectric generator, we have $\dot W$, $\dot Q_{L}>0$, and hence the condition for refrigeration, i.e., $\dot Q_{R}>0$, would imply $\dot W>\dot Q_{L}>0$ and a violation of the second law of thermodynamics. In fact, a thermodynamic generator cannot produce a power ($\dot W$) greater than the heat current taken from the hot reservoir ($\dot Q_{L}$). Hence, a thermodynamical machine can operate either as an engine or as a cooler.

Thus, a voltage biased asymmetric junction between two superconductors ($r\neq 1$) can behave either as a refrigerator or as a thermoelectric generator, depending on the temperature of the two electrodes $T_{ L},T_{ R}$. The competition of these two effects in a S'IS junction for a voltage bias $V=V_{ p}$ is displayed in Fig.~\ref{Fig4}, for $r=0.3$ (panel a) and for $r=0.75$ (panel b). The red areas denote the \textit{nonlinear-in-bias} thermoelectric regions $\dot W(V_{ p})>0$, whereas the blue areas give the cooling regions $\dot Q_{ R}(V_{ p})>0$. The dashed lines set the equal temperature contours $T_{ L}=T_{ R}$, and the vertical solid lines give the  thresholds $T_{ R}=T_{ c,R}=r T_{ c,L}$ (we recall that $r=T_{c,R}/T_{ c,L}=\Delta_{ 0,R}/\Delta_{ 0,L}$ for BCS superconductors). 

Let's focus first on $T_{ L}<T_{ R}$. Note that it is possible to remove the heat from the lower gap superconductor ($\dot Q_{ R}>0$), but necessarily there is no thermoelectricity, in agreement with the previous discussion. Note that this mechanism cannot be properly defined as cooling, since the heat is removed by the hotter electrode (sometimes called heat pump). However, this mechanism still relies on the existence of the larger superconducting gap. 

Consider now $T_{ L}\geq T_{ R}$. In this case, for a given value of $T_{ R}<T_{ c,R}$, the junction behaves as a refrigerator as long as the temperature difference is smaller than a threshold value. For larger values of $T_{ L}$ the junction is first dissipative and then it shows a thermoelectric generation for sufficiently high temperature gradients (roughly given by $T_{L}>T_{R}/r$, see red dashed lines). This progression from thermoelectricity towards cooling passing by a dissipative behaviour may remind the standard behaviour of the linear thermoelectricity~\cite{Benenti2017}. Anyway here there is a crucial difference. Namely, the parameter that control the transition from the cooling to the thermoelectricity is the temperature difference rather than the voltage bias. Furthermore, the thermoelectricity eventually disappears at large values of $T_{ L}$ where $\Delta_{ L}(T_{ L})<\Delta_{ R}(T_{ R})$ (solid red curves in Figs.~\ref{Fig5}a-b).
The plots also show that the correspondence between the thermoelectric effect and the evaporative cooling have some limitations. In fact, for $T_{ R}\geq T_{ c,R}$, i.e., when the smallest gap supercondutor is in the normal state, the evaporative cooling may be still achievable (see $T_{ R}>0.3T_{ L}$ in Fig.~\ref{Fig4}a), whereas the thermoelectric effect requires a monotonically decreasing DOS. This is guaranteed in our system only when the right electrode is in the superconducting state. 
\section{Heat Engine}
In the previous section, we discussed the theoretical features of the thermoelectric effect in a S'IS junction. In this section, we discuss the design of a heat engine based on this effect, for materials and a geometry which are experimentally feasible with standard nanofabrication techniques. Since we are interested in phenomena which require a temperature difference for nanoscale tunnel junctions, it is convenient to work with superconductors whose critical temperature is of order $1$ K, such as aluminum (Al), with a bulk critical temperature $T_{ c,Al}^{\rm bulk}=1.2$~K. In fact, at sub-Kelvin temperatures, the electron-phonon coupling is quite weak and hence it is possible to raise the quasiparticle temperature well above the bath temperature $T_{\rm bath}$, which typically represents the temperature of the phonons in the electrodes~\cite{GiazottoRMP,Muhonen2012}.
This condition is known in the literature as \textit{quasi-equilibrium} regime~\cite{GiazottoRMP,Muhonen2012}, since both the quasiparticles temperature and the phonons temperature are well-defined, but they can be different. 
The validity of this regime has been demonstrated and investigated in several experiments which involves the electronic temperature cooling or the coherent control of the heat currents in superconducting tunnel junctions~\cite{Muhonen2012,FornieriReview,Hwang2020}. 
In particular, Al is an optimal choice for our purposes, due to the excellent control of the quality in aluminum-oxide based tunnel junctions~\cite{Gurvitch1983}. The latter is an important requirement in order to suppress any unwanted Josephson contribution. The condition $r\neq 1$ can be achieved in thin bi-layers where aluminum is used in combination with other materials, such as a superconductor with lower gap as titanium (Ti)~\cite{Lolli2016} or a normal metal as copper (Cu)~\cite{FornieriTimossi}. More precisely, the gap is reduced with respect to a fully aluminum based structure due to inverse proximity effect~\cite{DEGENNESRMP}. In this section (unless explicitly stated), we consider a thin aluminum film for S' with $T_{\rm c,Al}^{\rm film}\sim1.32$ K and gap $\Delta_{0,S'}=200\mu$eV and an Al-Cu bilayer with $\Delta_{0,S}=0.3\Delta_{0,S'}\sim67\mu$eV and $T_{\rm c,by}=0.3 T_{\rm c,Al}^{\rm film}\sim 0.44$ K.

\begin{figure}
   \includegraphics[width=0.5\textwidth]{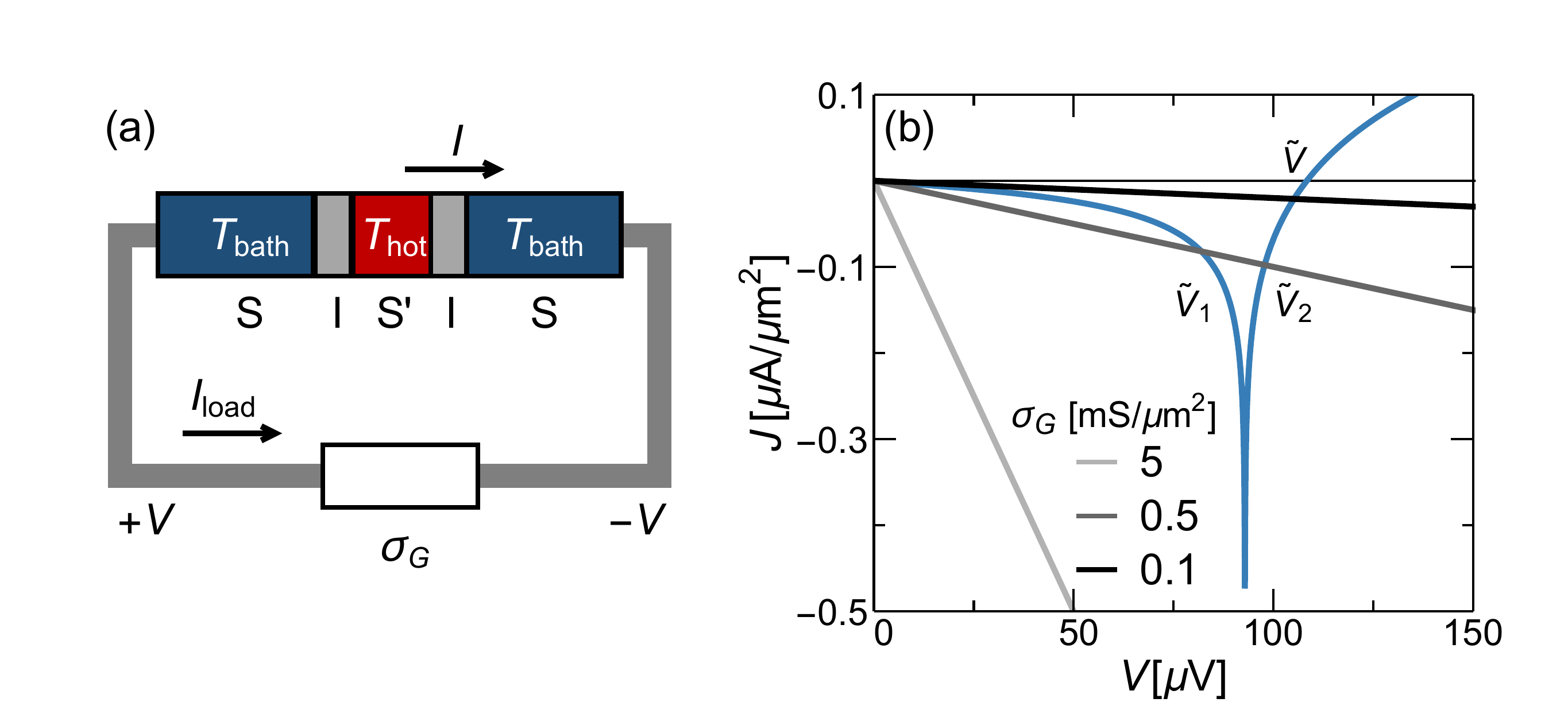}
    \caption{a) Scheme of the heat engine based on the thermoelectric effect in a superconducting junction. The system is composed of two superconducting junctions connected back to back (SIS'IS). A temperature gradient is applied between the central superconductor (red) and the two lateral superconductors (blue), i.e, $T_{\rm S'}=T_{\rm hot}>T_{\rm bath}=T_{\rm S}$. Under proper conditions, the system  spontaneously develops a voltage bias $2V$ across the resistor ($V$ is the voltage drop across each S'IS junction), and hence a thermoelectric current, which releases power to the load. b) Graphical solution of Eq.~\ref{eq:self-current-circuit} for different values of unit surface conductance $\sigma_{ G}$. The thermoelectric voltage $V$ is given by the crossing points of the $I(V)$ characteristic (blue curve) and the load lines (grayscale). Parameters: $T_{\rm hot}=1$ K, $T_{\rm cold}=0.1$ K.}
    \label{Fig5}
\end{figure}

The scheme of the heat engine is pictured in Fig.~\ref{Fig5}a. The system consists of the series of two S'IS junction connected back to back, in a SIS'IS configuration, in parallel with a load of conductance $G$. The central element (red) is the larger gap superconductor (Al), whereas the lateral superconductors have a smaller gap (Al-Cu bilayers). The lateral superconductors are strongly coupled to the phonon bath thanks to their large volume, thus the quasiparticle temperature in the S layers nominally resides at $T_{\rm bath}$. We assume that the electronic temperature of the S' island $T_{\rm hot}$ is instead raised above $T_{\rm bath}$, typically using other superconducting or normal metal tunnel junctions as heaters~\cite{FornieriTimossi,FornieriReview}. In this configuration, the thermoelectric contributions of the two S'IS junctions add. Indeed, in the presence of a thermal gradient between the central superconductor (hot) and the lateral superconductors (cold), a thermoelectric voltage develops across the whole structure (see the discussion below).
As a consequence of thermoelectricity, a voltage $V_{\rm load}$ develops across the load, and a current $I_{\rm load}=G V_{\rm load}$ flows through the structure. Thus, a power $\dot W=I_{\rm load}V_{\rm load}$ is delivered to the load. For convenience, we consider a symmetric structure SIS'IS junction (see Fig.~\ref{Fig5}a). Note that the crucial constraint in this configuration is represented by the current conservation in the circuit, which guarantees that the voltage drops of the two junctions add. Due to symmetry, the voltage drop across the load is $V_{\rm load}=2V$, where $V$ is the voltage drop across each S'IS junction.  Hence, the use of a SIS'IS structure produces a doubled thermoelectric voltage with respect to the single junction. Moreover, this symmetric configuration is also convenient in terms of the shadow mask evaporation, which is the common fabrication technique for high quality tunnel junctions based on Al, and has been exploited in several experiments~\cite{GiazottoRMP,Muhonen2012}. Finally, we note that in a fully symmetric structure, in the presence \emph{only} of a standard linear thermoelectric effect, this configuration \emph{would not} produce a finite difference between the two lateral leads, since the charge diffusion in the left and the right lead would cancel out due to the opposite temperature gradients. This fact demonstrates the unique features of the nonlinear thermoelectricity in the system here described.

In order to compute the thermovoltage and $\dot W_{\rm load}$, one has to impose
the current conservation in the circuit, namely
\begin{equation}
I_{\rm load}=GV_{\rm load}=2GV=-I(V,T_{\rm hot},T_{\rm bath})
\label{eq:self-current-circuit}
\end{equation}
 and to solve it self-consistently in $V$. 
 Due to the EH symmetry, this equation admits always the trivial solution $V=0$, where the current $I_{\rm load}$ (and hence the delivered power $\dot W_{\rm load}$) is zero. In the presence of thermoelectricity, the junction displays an absolute negative conductance for biases below $V_{ S}$, and hence additional solutions with finite voltage $\tilde V\neq 0$ are possible (see Fig.~\ref{Fig5}b for an example). Due to EH symmetry, for each finite solution $V=\tilde V$ there is a correspondent solution $V=-\tilde V$, i.e., finite values solutions always come in pairs $\pm\tilde V$. Since the conductance $G_{ T}$ and hence relevant quantities such as the current $I$ and the thermoelectric power $\dot W$ are proportional to the surface $\mathcal A$ of the tunnel junction, we discuss their value for unit surface. In particular, we consider realistic tunnel junctions with specific barrier conductance of $\sigma_{T}=10$ mS$/\mu$m$^2$. Hence, it is convenient to introduce in the discussion a load conductance for unit area (defining $\sigma_{G}=G/ \mathcal A$), in order to express the figures of merit of the heat engine in a scale-invariant fashion. The absolute values are easily obtained by multiplying for a specific surface, such as 1$\mu$m$^2$.
 
 From a geometric view, the solutions of Eq.~\ref{eq:self-current-circuit} are the crossings of the current density characteristic $J(V)=I(V)/\mathcal A$ with a load line of negative slope $-2\sigma_{\rm G}$, as displayed in Fig.~\ref{Fig5}b for different values of $\sigma_{\rm G}$. In the plot, we set $T_{\rm hot}=1$K and $T_{\rm bath}=0.1$ K, so that both the \textit{linear-in-bias} and the \textit{nonlinear-in-bias} contributions to the thermoelectricity are present. This represents the typicality of the effect, as already discussed in the previous section.
 In this case, there are mainly three situations, related to the values  $\sigma_{ p}=|G_{p}|/\mathcal A\sim 5$ mS/$\mu$m$^2$ and $\sigma_{0}=|G_{\rm 0}|/\mathcal A\sim 0.37$ mS/$\mu$m$^2$: i) for $\sigma_{ G}>|\sigma_{ p}|/2$, there is no solution with $V\neq0$, ii) for $|\sigma_{\rm p}|/2>\sigma_{\rm G}>|\sigma_{\rm 0}|/2\sim 0.18$ mS/$\mu$m$^2$, there are two positive solutions $\tilde V_{1}<V_{p}$ and $\tilde V_{ 2}>V_{ p}$ (and hence a total of 5 solutions, due to EH symmetry), iii) $\sigma_{ G}<|\sigma_{\rm 0}|/2$ there are three solutions $V=0,\pm \tilde V$ (see Fig.~\ref{Fig5}b).  In this work, we will only focus on the solutions characterized by a positive slope of the $J(V)$  characteristic [either $\tilde V$ or $\tilde V_{2}$], which are stable independently by the details of the load circuit, such as the parasite capacitance and the self-inductance. In particular, the instability of the $V=0$ solution for $\sigma_{G}>|\sigma_{\rm peak}|/2$ can lead to an oscillatory behavior, which goes beyond the purpose of this work~\cite{MarchegianiNLTE}. 
 Since we discuss only the stationary and time independent solutions and we completely neglect those cases.
 In summary, for $\sigma_{\rm G}<2|\sigma_{ p}|/2$, a thermoelectric voltage $V$ develops across each S'IS junction and the system provides a thermoelectric power density $\dot{W}/\mathcal A=I_{\rm load}V_{\rm load}/\mathcal A=4\sigma_{\rm G}V^2$.
 
 \subsection{Load dependence of power and efficiency}
 Here, we discuss the thermoelectric power density and the corresponding thermodynamical efficiency as a function of the load conductivity $\sigma_{\rm G}$. Note that, for a given thermoelectric configuration, characterized by the parameters $T_{\rm hot},T_{\rm bath},r$, the power and the efficiency are zero either for $\sigma_{\rm G}=0$ since $I_{\rm load}=0$ and for $\sigma_{\rm G}>|\sigma_{\rm p}|/2$, where $V_{\rm load}=0$.
 
\begin{figure}
    \includegraphics[width=0.5\textwidth]{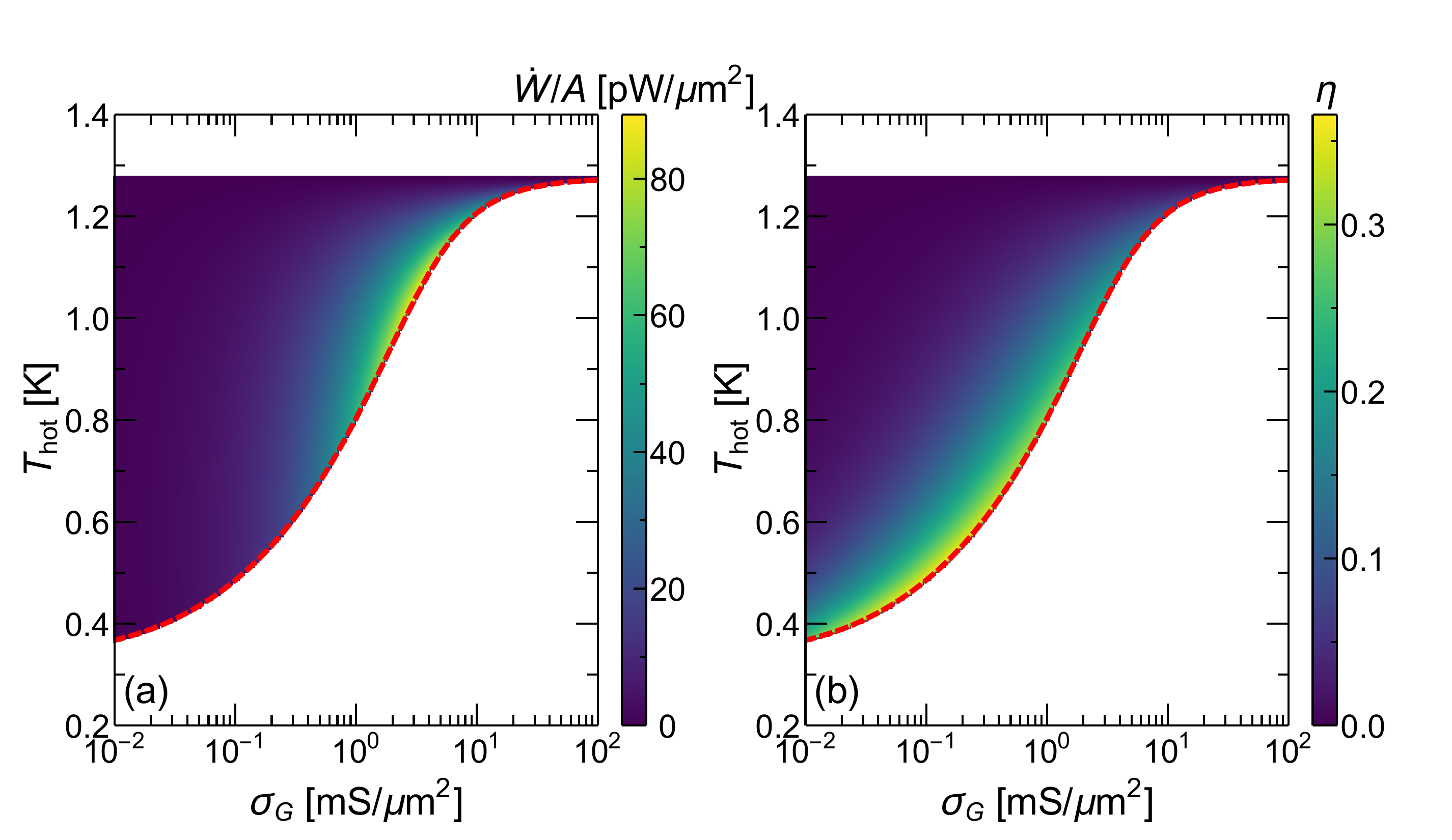}
    \caption{Density plot of the thermoelectric power density (a) and the thermodynamical efficiency (b) as a function of the temperature of the S' island and the specific conductance of the load. The white regions correspond to a zero value. The contours $\sigma_{G}=\sigma_{ p}(T_L)/2$ are drawn with red dashed curves.} 
    \label{Fig6}
\end{figure}

 Figure~\ref{Fig6}a displays the density plot of the thermoelectric power density $\dot{W}/\mathcal A$ as a function of the specific conductance of the load $\sigma_{ G}$ and the temperature of the hot electrode $T_{\rm hot}$ for $T_{\rm cold}=100$ mK (and so $\Delta_{S}(T_{\rm cold})\simeq \Delta_{ 0,S}\sim 67\mu$eV). The corresponding thermoelectric efficiency $\eta=\dot{W}/(2\dot Q_{\rm hot})$~\footnote{The factor 2 takes into account the presence of the two junctions.} is displayed in Fig.~\ref{Fig6}b. 
 In both the plots, there are two white regions where the thermoelectric power is absent (Fig.~\ref{Fig6}a) and the efficiency is consequently zero (Fig.~\ref{Fig6}b). These areas correspond to: i) $T_{\rm  hot}\geq 1.27$ K, where $\Delta_{ S'}<\Delta_{ S}$; ii) large values of $\sigma_{G}$, where Eq.~\ref{eq:self-current-circuit} has only the zero-voltage solution.
 For a given value of $T_{\rm  hot}$, both $\dot W/\mathcal A$ and $\eta$ are maximum for $\sigma_{G}\lesssim |\sigma_{ p}(T_{\rm  hot})|/\mathcal S$ (red dashed curves), and they worsen by reducing $\sigma_{G}$. At low values of $\sigma_{G}$, the systems works as a heat engine over a large range in $T_{\rm  hot}$, but the power and, for large temperature gradients, the efficiency are typically reduced. At higher values of $\sigma_{G}$ one finds increased performance but a reduced operative range in terms of $T_{\rm  hot}$. Thus, there is a trade off between the thermoelectric performance and the operative temperature range. The maximum power density reads $\dot W_{\rm max}/\mathcal S\sim 2\times 0.11 \sigma_{ T}\Delta_{0,S'}^2/e^2\sim 88$ pW/$\mu$m$^2$ and the maximum efficiency is roughly $\eta_{\rm max}=0.36$.

\subsection{Seebeck voltage and nonlinear Seebeck coefficient}
\begin{figure}
    \includegraphics[width=0.5\textwidth]{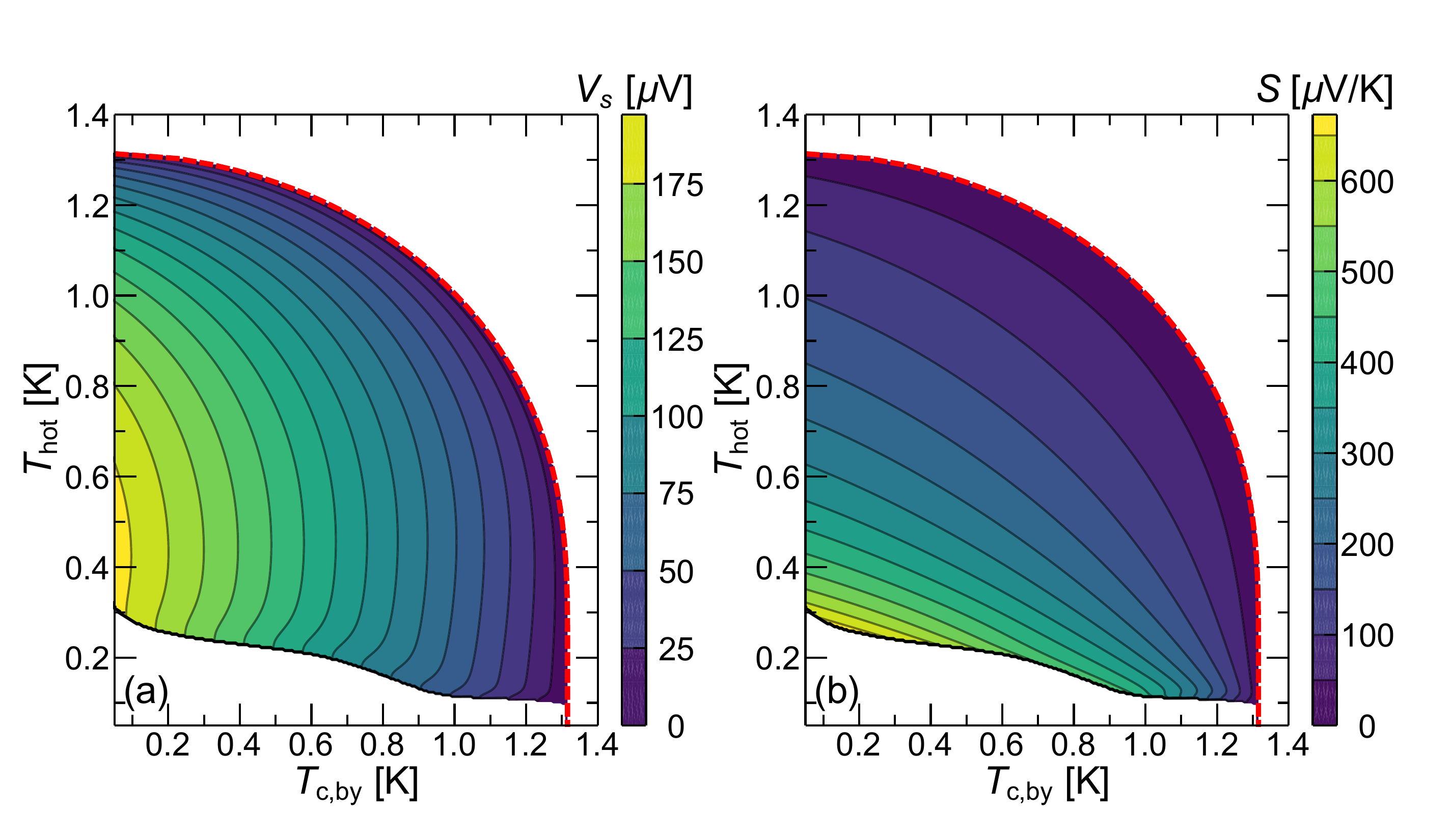}
    \caption{Contour plot of the Seebeck voltage (a) and the thermodynamical efficiency (b) as a function of the temperature of the S' island and the critical temperature of the bilayer S. The white regions correspond to a zero value. The red lines gives the constraint $\Delta_{S'}(T_{\rm hot})=\Delta_{S}=1.764 k_B T_{\rm c,by}$.}    \label{Fig7}
\end{figure}
In a open circuit configuration, i.e., in the limit $\sigma_{G}\to 0$, the current $I_{\rm load}=0$ and hence the thermoelectric power is zero. In this case, the thermoelectric effect purely manifests as a voltage signal across the load $V_{\rm load}=2V_{s}$, where $V_{s}$ is the Seebeck voltage introduced in Sec.~\ref{sec:model}. Figure~\ref{Fig7}a displays the contour plot of the Seebeck voltage $V_{s}$ as a function of the critical temperature of the bilayer $T_{c,by}$ and the temperature of the hot electrode $T_{\rm hot}$ for $T_{\rm bath}\ll T_{c,by}$, assuming to keep fixed the critical temperature of the hot terminal $T_{c,S'}\sim 1.32$K. For a given $T_{\rm hot}$, $V_{s}$ is monotonically decreasing with $T_{c,by}$ and it is zero (white) when $T_{\rm c,by}$ is larger than a threshold value, i.e., $T_{\rm c,by}\geq \Delta_{\rm S'}(T_{\rm hot})/(1.764 k_B)$. The white region at low values of $T_{\rm hot}$ is related to the finite value of the Dynes parameters (see the discussion in the next subsection). Note that the maximum Seebeck voltage is roughly given by $\Delta_{\rm 0,S'}/e\sim 200\mu$V. A similar behavior apply to the corresponding nonlinear Seebeck coefficient, defined as $\mathcal S=V_{\rm s}/\Delta T$, with $\Delta T=T_{\rm hot}-T_{\rm bath}$ and displayed in Fig.~\ref{Fig7}b. Notably, $\mathcal S$ has a value of hundreds of $\mu$V/K over large temperature ranges and can reach a value as large as 650 $\mu$V/K for $T_{\rm c,by}\sim 0.2$K and $T_{\rm hot}\sim 0.3$K.  
\subsection{Effect of nonidealities}
\begin{figure}
    \includegraphics[width=\columnwidth]{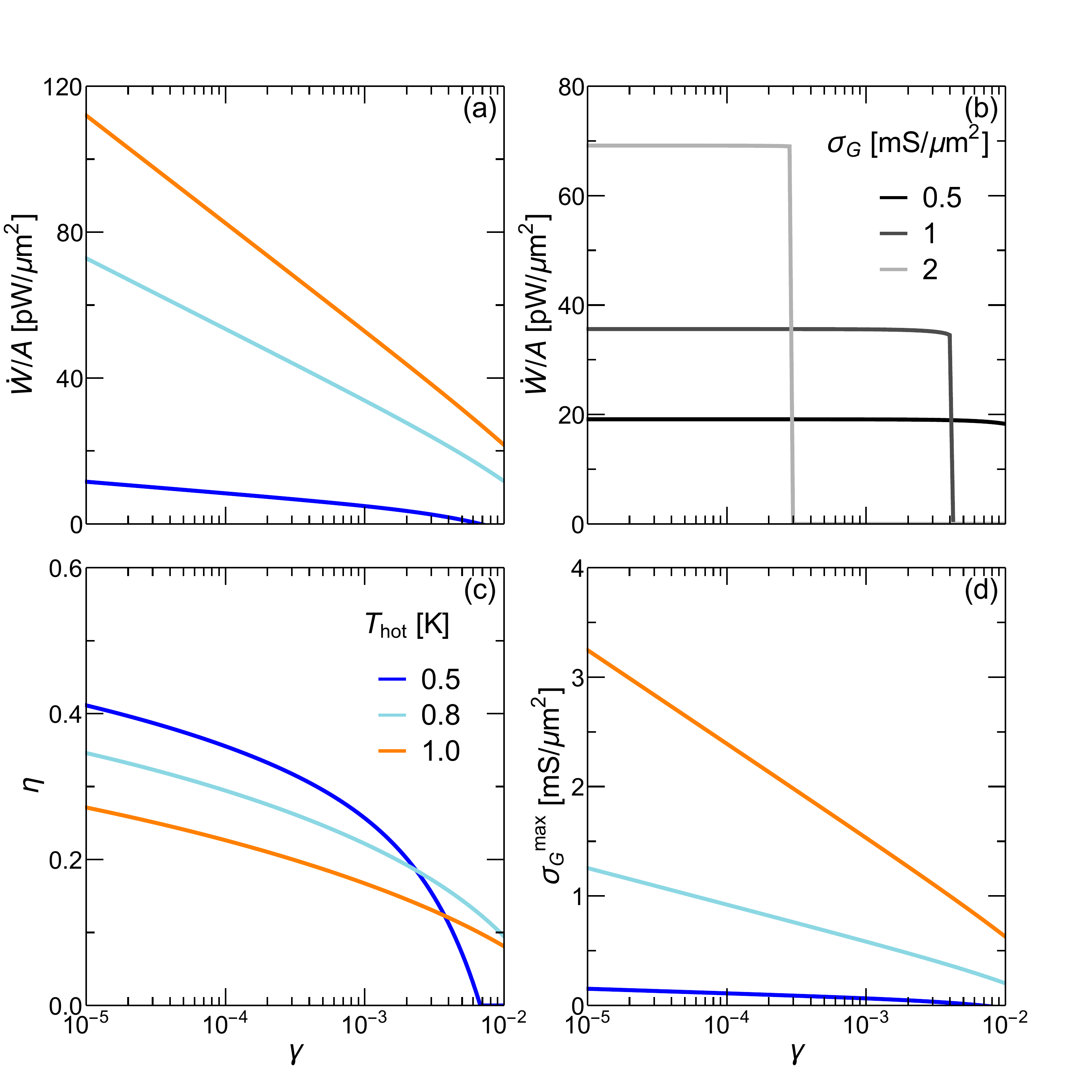}
    \caption{Impact of the rescaled Dynes parameter $\gamma=\Gamma_S'/\Delta_{\rm 0,S'}$. $\gamma$-evolution of the matching peak value of the power density (a) and the thermodynamical efficiency (c) for different values of $T_{\rm hot}$. (b) Power density vs $\gamma$ for $T_{\rm hot}=$ and different values of $\sigma_{\rm G}$. (d) Maximum load supported by the heat engine vs $\gamma$ for the same values of $T_{\rm hot}$ as in panels (a) and (c). Parameters: $T_{\rm c,S'}\sim 1.32$K and $T_{\rm c,by}=0.3T_{\rm c,S'}\sim 0.44$ K.}
    \label{Fig8}
\end{figure}
Here, we want to characterize the impact of the main source of non-ideality in our model, namely the Dynes parameters $\Gamma_{\rm \alpha}$. In fact, these parameters characterize either the finite number of states at subgap energies of the BCS superconducting DOS and the smoothing of the peaks in superconducting DOS. As a consequence, the current at $V_p$ and hence relevant quantities such as the thermoelectric power are reduced. Differently from the rest of this work, here we introduce a dimensionless parameter $\gamma$ and we consider equal values for the Dynes parameters $\Gamma_{S'}=\Gamma_{S}=\gamma \Delta_{\rm 0,S'}=\max(\Gamma_{S}/\Gamma_{S'})$ in order to overestimate the worsening effect. In the plots, we set $T_{\rm bath}=100$ mK.
First, we consider the quantities where the variation of the Dynes parameter is expected to impact in a stronger way, namely the thermoelectric power and the efficiency at the matching peak singularity $V_{p}=[\Delta_{S'}-\Delta_{S}]/e$. These quantities are displayed for different values of $T_{\rm hot}$ in Fig.~\ref{Fig8}a and Fig.~\ref{Fig8}c, respectively. Note that both $\dot W/\mathcal A$ and $\eta$ decreases monotonically by increasing $\gamma$, as expected. Interestingly they are typically reduced only by a factor 3-4 under orders of magnitude in $\gamma$ from $10^{-5}$ to $10^{-2}$, showing that the thermoelectric effect is quite robust against $\gamma$. However, a large value of $\gamma$ may suppress completely the thermoelectric effect when the thermoelectric power is quite low, as shown by the curves corresponding to the lowest temperature (blue), for $\gamma\geq 7\times10^{-3}$.

The impact of the Dynes parameter is even less relevant if the system is not biased with a voltage equal to the matching singularity peak. This feature is shown in Fig.~\ref{Fig8}b, where the thermoelectric power density, obtain through the self-consistent solution of Eq.~\ref{eq:self-current-circuit}, is displayed for $T_{\rm hot}=1$K and different values of $\sigma_{\rm G}$. In particular, the power is roughly constant up to a threshold value, depending on $\sigma_{\rm G}$, where the thermoelectric effect goes to zero. This feature can be understood by inquiring the graphical solution of Eq.~\ref{eq:self-current-circuit} displayed in Fig.~\ref{Fig5}b. In particular, we recall that Eq.~\ref{eq:self-current-circuit} have no finite solution for $\sigma_{G}>\sigma_{ p}/2$. Upon increasing $\gamma$, the current at the matching peak singularity decreases while $V_{\rm p}$ is fixed but again a big variation of $\gamma$ affect with a small multiplicative factor. As a consequence, the absolute value of $\sigma_{ p}$ is reduced as well and a large value of $\gamma$ may produce a situation where there are no crossing for $V\neq 0$. In this context, Fig.~\ref{Fig8}d displays the maximum value of the specific conductance of the load supported by the thermoelectric generator for different values of $T_{\rm hot}$. As discussed above, this value monotonically decreases with $\gamma$.
\section{Conclusions}
In summary, we have given an extended discussion of the nonlinear thermoelectric effect recently predicted in tunnel junctions between two different BCS superconductors~\cite{MarchegianiNLTE}. The thermoelectric generation occurs when the temperature difference is larger than a threshold value and the hot electrode has the largest gap. We focused on two region: the \textit{linear-in-bias} contribution, characterized by a negative differential conductance at $V=0$ and the \textit{nonlinear-in-bias} contribution, where the thermoelectric performance is optimal. We argued that this effect is somewhat complementary to the evaporative cooling in superconducting junctions due to the presence of the gap. However, the thermoelectric generation has tighter requirements, since it requires also a locally monotonically decreasing DOS in the cold electrode. Finally, we presented a design study for an experiment involving a heat engine based on the thermoelectric effect for an Al-based structure. We characterized the main thermoelectric figures of merit, predicting a power density up to 88 pW/$\mu$m$^2$ and efficiencies up to 40\%. Correspondingly, we show that one can observe a Seebeck potential of the order of $200\mu$V and a nonlinear Seebeck coefficient up to $650\mu$V/K for realistic parameter values. Finally, we discussed how the performance is weakly affected by non-idealities such as the Dynes parameter. The engine can be experimentally realised with current state of the art nanotechnology. The successful confirmation of the discussed phenomenology would potentially trigger further research on the same thermoelectric mechanism in other physical systems.

\begin{acknowledgments}
We acknowledge the Horizon research and innovation programme under grant agreement No. 800923 (SUPERTED) for partial financial support. A.B. acknowledges the CNR-CONICET cooperation program "Energy conversion in quantum nanoscale hybrid devices", the SNS-WIS jointlab QUANTRA funded by the Italian Ministry of Foreign Affairs and International Cooperation, and the Royal Society through
the International Exchanges between the UK and Italy (Grants No. IES R3 170054 and IEC R2 192166).
\end{acknowledgments}


\begin{thebibliography}{61}%
	\makeatletter
	\providecommand \@ifxundefined [1]{%
		\@ifx{#1\undefined}
	}%
	\providecommand \@ifnum [1]{%
		\ifnum #1\expandafter \@firstoftwo
		\else \expandafter \@secondoftwo
		\fi
	}%
	\providecommand \@ifx [1]{%
		\ifx #1\expandafter \@firstoftwo
		\else \expandafter \@secondoftwo
		\fi
	}%
	\providecommand \natexlab [1]{#1}%
	\providecommand \enquote  [1]{``#1''}%
	\providecommand \bibnamefont  [1]{#1}%
	\providecommand \bibfnamefont [1]{#1}%
	\providecommand \citenamefont [1]{#1}%
	\providecommand \href@noop [0]{\@secondoftwo}%
	\providecommand \href [0]{\begingroup \@sanitize@url \@href}%
	\providecommand \@href[1]{\@@startlink{#1}\@@href}%
	\providecommand \@@href[1]{\endgroup#1\@@endlink}%
	\providecommand \@sanitize@url [0]{\catcode `\\12\catcode `\$12\catcode
		`\&12\catcode `\#12\catcode `\^12\catcode `\_12\catcode `\%12\relax}%
	\providecommand \@@startlink[1]{}%
	\providecommand \@@endlink[0]{}%
	\providecommand \url  [0]{\begingroup\@sanitize@url \@url }%
	\providecommand \@url [1]{\endgroup\@href {#1}{\urlprefix }}%
	\providecommand \urlprefix  [0]{URL }%
	\providecommand \Eprint [0]{\href }%
	\providecommand \doibase [0]{https://doi.org/}%
	\providecommand \selectlanguage [0]{\@gobble}%
	\providecommand \bibinfo  [0]{\@secondoftwo}%
	\providecommand \bibfield  [0]{\@secondoftwo}%
	\providecommand \translation [1]{[#1]}%
	\providecommand \BibitemOpen [0]{}%
	\providecommand \bibitemStop [0]{}%
	\providecommand \bibitemNoStop [0]{.\EOS\space}%
	\providecommand \EOS [0]{\spacefactor3000\relax}%
	\providecommand \BibitemShut  [1]{\csname bibitem#1\endcsname}%
	\let\auto@bib@innerbib\@empty
	%</preamble>
	\bibitem [{\citenamefont {Benenti}\ \emph {et~al.}(2017)\citenamefont
		{Benenti}, \citenamefont {Casati}, \citenamefont {Saito},\ and\ \citenamefont
		{Whitney}}]{Benenti2017}%
	\BibitemOpen
	\bibfield  {author} {\bibinfo {author} {\bibfnamefont {G.}~\bibnamefont
			{Benenti}}, \bibinfo {author} {\bibfnamefont {G.}~\bibnamefont {Casati}},
		\bibinfo {author} {\bibfnamefont {K.}~\bibnamefont {Saito}},\ and\ \bibinfo
		{author} {\bibfnamefont {R.}~\bibnamefont {Whitney}},\ }\bibfield  {title}
	{\bibinfo {title} {Fundamental aspects of steady-state conversion of heat to
			work at the nanoscale},\ }\href
	{https://doi.org/https://doi.org/10.1016/j.physrep.2017.05.008} {\bibfield
		{journal} {\bibinfo  {journal} {Phys. Rep.}\ }\textbf {\bibinfo {volume}
			{694}},\ \bibinfo {pages} {1 } (\bibinfo {year} {2017})}\BibitemShut
	{NoStop}%
	\bibitem [{\citenamefont {Pop}\ \emph {et~al.}(2006)\citenamefont {Pop},
		\citenamefont {Sinha},\ and\ \citenamefont {Goodson}}]{Goodson2006}%
	\BibitemOpen
	\bibfield  {author} {\bibinfo {author} {\bibfnamefont {E.}~\bibnamefont
			{Pop}}, \bibinfo {author} {\bibfnamefont {S.}~\bibnamefont {Sinha}},\ and\
		\bibinfo {author} {\bibfnamefont {K.~E.}\ \bibnamefont {Goodson}},\
	}\bibfield  {title} {\bibinfo {title} {Heat generation and transport in
			nanometer-scale transistors},\ }\href
	{https://doi.org/10.1109/JPROC.2006.879794} {\bibfield  {journal} {\bibinfo
			{journal} {Proceedings of the IEEE}\ }\textbf {\bibinfo {volume} {94}},\
		\bibinfo {pages} {1587} (\bibinfo {year} {2006})}\BibitemShut {NoStop}%
	\bibitem [{\citenamefont {Giazotto}\ \emph {et~al.}(2006)\citenamefont
		{Giazotto}, \citenamefont {Heikkil{\"{a}}}, \citenamefont {Luukanen},
		\citenamefont {Savin},\ and\ \citenamefont {Pekola}}]{GiazottoRMP}%
	\BibitemOpen
	\bibfield  {author} {\bibinfo {author} {\bibfnamefont {F.}~\bibnamefont
			{Giazotto}}, \bibinfo {author} {\bibfnamefont {T.~T.}\ \bibnamefont
			{Heikkil{\"{a}}}}, \bibinfo {author} {\bibfnamefont {A.}~\bibnamefont
			{Luukanen}}, \bibinfo {author} {\bibfnamefont {A.~M.}\ \bibnamefont
			{Savin}},\ and\ \bibinfo {author} {\bibfnamefont {J.~P.}\ \bibnamefont
			{Pekola}},\ }\bibfield  {title} {\bibinfo {title} {Opportunities for
			mesoscopics in thermometry and refrigeration: Physics and applications},\
	}\href {https://doi.org/10.1103/RevModPhys.78.217} {\bibfield  {journal}
		{\bibinfo  {journal} {Rev. Mod. Phys.}\ }\textbf {\bibinfo {volume} {78}},\
		\bibinfo {pages} {217} (\bibinfo {year} {2006})}\BibitemShut {NoStop}%
	\bibitem [{\citenamefont {Muhonen}\ \emph {et~al.}(2012)\citenamefont
		{Muhonen}, \citenamefont {Meschke},\ and\ \citenamefont
		{Pekola}}]{Muhonen2012}%
	\BibitemOpen
	\bibfield  {author} {\bibinfo {author} {\bibfnamefont {J.~T.}\ \bibnamefont
			{Muhonen}}, \bibinfo {author} {\bibfnamefont {M.}~\bibnamefont {Meschke}},\
		and\ \bibinfo {author} {\bibfnamefont {J.~P.}\ \bibnamefont {Pekola}},\
	}\bibfield  {title} {\bibinfo {title} {Micrometre-scale refrigerators},\
	}\href {https://doi.org/10.1088/0034-4885/75/4/046501} {\bibfield  {journal}
		{\bibinfo  {journal} {Rep. Prog. Phys.}\ }\textbf {\bibinfo {volume} {75}},\
		\bibinfo {pages} {046501} (\bibinfo {year} {2012})}\BibitemShut {NoStop}%
	\bibitem [{\citenamefont {Dubi}\ and\ \citenamefont
		{Di~Ventra}(2011)}]{DiVentraRMP}%
	\BibitemOpen
	\bibfield  {author} {\bibinfo {author} {\bibfnamefont {Y.}~\bibnamefont
			{Dubi}}\ and\ \bibinfo {author} {\bibfnamefont {M.}~\bibnamefont
			{Di~Ventra}},\ }\bibfield  {title} {\bibinfo {title} {Colloquium: Heat flow
			and thermoelectricity in atomic and molecular junctions},\ }\href
	{https://doi.org/10.1103/RevModPhys.83.131} {\bibfield  {journal} {\bibinfo
			{journal} {Rev. Mod. Phys.}\ }\textbf {\bibinfo {volume} {83}},\ \bibinfo
		{pages} {131} (\bibinfo {year} {2011})}\BibitemShut {NoStop}%
	\bibitem [{\citenamefont {Kosloff}(2013)}]{KosloffEntropy}%
	\BibitemOpen
	\bibfield  {author} {\bibinfo {author} {\bibfnamefont {R.}~\bibnamefont
			{Kosloff}},\ }\bibfield  {title} {\bibinfo {title} {Quantum thermodynamics: A
			dynamical viewpoint},\ }\href@noop {} {\bibfield  {journal} {\bibinfo
			{journal} {Entropy}\ }\textbf {\bibinfo {volume} {15}},\ \bibinfo {pages}
		{2100} (\bibinfo {year} {2013})}\BibitemShut {NoStop}%
	\bibitem [{\citenamefont {Cahill}\ \emph {et~al.}(2003)\citenamefont {Cahill},
		\citenamefont {Ford}, \citenamefont {Goodson}, \citenamefont {Mahan},
		\citenamefont {Majumdar}, \citenamefont {Maris}, \citenamefont {Merlin},\
		and\ \citenamefont {Phillpot}}]{CahillJAP2003}%
	\BibitemOpen
	\bibfield  {author} {\bibinfo {author} {\bibfnamefont {D.~G.}\ \bibnamefont
			{Cahill}}, \bibinfo {author} {\bibfnamefont {W.~K.}\ \bibnamefont {Ford}},
		\bibinfo {author} {\bibfnamefont {K.~E.}\ \bibnamefont {Goodson}}, \bibinfo
		{author} {\bibfnamefont {G.~D.}\ \bibnamefont {Mahan}}, \bibinfo {author}
		{\bibfnamefont {A.}~\bibnamefont {Majumdar}}, \bibinfo {author}
		{\bibfnamefont {H.~J.}\ \bibnamefont {Maris}}, \bibinfo {author}
		{\bibfnamefont {R.}~\bibnamefont {Merlin}},\ and\ \bibinfo {author}
		{\bibfnamefont {S.~R.}\ \bibnamefont {Phillpot}},\ }\bibfield  {title}
	{\bibinfo {title} {Nanoscale thermal transport},\ }\href
	{https://doi.org/10.1063/1.1524305} {\bibfield  {journal} {\bibinfo
			{journal} {J. Appl. Phys.}\ }\textbf {\bibinfo {volume} {93}},\ \bibinfo
		{pages} {793} (\bibinfo {year} {2003})}\BibitemShut {NoStop}%
	\bibitem [{\citenamefont {Bergfield}\ and\ \citenamefont
		{Ratner}(2013)}]{BergfieldMolecular}%
	\BibitemOpen
	\bibfield  {author} {\bibinfo {author} {\bibfnamefont {J.~P.}\ \bibnamefont
			{Bergfield}}\ and\ \bibinfo {author} {\bibfnamefont {M.~A.}\ \bibnamefont
			{Ratner}},\ }\bibfield  {title} {\bibinfo {title} {Forty years of molecular
			electronics: Non-equilibrium heat and charge transport at the nanoscale},\
	}\href {https://doi.org/10.1002/pssb.201350048} {\bibfield  {journal}
		{\bibinfo  {journal} {Phys. Status Solidi B}\ }\textbf {\bibinfo {volume}
			{250}},\ \bibinfo {pages} {2249} (\bibinfo {year} {2013})}\BibitemShut
	{NoStop}%
	\bibitem [{\citenamefont {{Wang, J.-S.}}\ \emph {et~al.}(2008)\citenamefont
		{{Wang, J.-S.}}, \citenamefont {{Wang, J.}},\ and\ \citenamefont {{L\"u, J.
				T.}}}]{WangEPJB}%
	\BibitemOpen
	\bibfield  {author} {\bibinfo {author} {\bibnamefont {{Wang, J.-S.}}},
		\bibinfo {author} {\bibnamefont {{Wang, J.}}},\ and\ \bibinfo {author}
		{\bibnamefont {{L\"u, J. T.}}},\ }\bibfield  {title} {\bibinfo {title}
		{Quantum thermal transport in nanostructures},\ }\href
	{https://doi.org/10.1140/epjb/e2008-00195-8} {\bibfield  {journal} {\bibinfo
			{journal} {Eur. Phys. J. B}\ }\textbf {\bibinfo {volume} {62}},\ \bibinfo
		{pages} {381} (\bibinfo {year} {2008})}\BibitemShut {NoStop}%
	\bibitem [{\citenamefont {Pop}(2010)}]{Pop2010}%
	\BibitemOpen
	\bibfield  {author} {\bibinfo {author} {\bibfnamefont {E.}~\bibnamefont
			{Pop}},\ }\bibfield  {title} {\bibinfo {title} {Energy dissipation and
			transport in nanoscale devices},\ }\href
	{https://doi.org/10.1007/s12274-010-1019-z} {\bibfield  {journal} {\bibinfo
			{journal} {Nano Res.}\ }\textbf {\bibinfo {volume} {3}},\ \bibinfo {pages}
		{147} (\bibinfo {year} {2010})}\BibitemShut {NoStop}%
	\bibitem [{\citenamefont {Chowdhury}\ \emph {et~al.}(2009)\citenamefont
		{Chowdhury}, \citenamefont {Prasher}, \citenamefont {Lofgreen}, \citenamefont
		{Chrysler}, \citenamefont {Narasimhan}, \citenamefont {Mahajan},
		\citenamefont {Koester}, \citenamefont {Alley},\ and\ \citenamefont
		{Venkatasubramanian}}]{Chowdhury2009}%
	\BibitemOpen
	\bibfield  {author} {\bibinfo {author} {\bibfnamefont {I.}~\bibnamefont
			{Chowdhury}}, \bibinfo {author} {\bibfnamefont {R.}~\bibnamefont {Prasher}},
		\bibinfo {author} {\bibfnamefont {K.}~\bibnamefont {Lofgreen}}, \bibinfo
		{author} {\bibfnamefont {G.}~\bibnamefont {Chrysler}}, \bibinfo {author}
		{\bibfnamefont {S.}~\bibnamefont {Narasimhan}}, \bibinfo {author}
		{\bibfnamefont {R.}~\bibnamefont {Mahajan}}, \bibinfo {author} {\bibfnamefont
			{D.}~\bibnamefont {Koester}}, \bibinfo {author} {\bibfnamefont
			{R.}~\bibnamefont {Alley}},\ and\ \bibinfo {author} {\bibfnamefont
			{R.}~\bibnamefont {Venkatasubramanian}},\ }\bibfield  {title} {\bibinfo
		{title} {On-chip cooling by superlattice-based thin-film thermoelectrics},\
	}\href {https://doi.org/10.1038/nnano.2008.417} {\bibfield  {journal}
		{\bibinfo  {journal} {Nat. Nanotech.}\ }\textbf {\bibinfo {volume} {4}},\
		\bibinfo {pages} {235} (\bibinfo {year} {2009})}\BibitemShut {NoStop}%
	\bibitem [{\citenamefont {Bradley}\ \emph {et~al.}(2017)\citenamefont
		{Bradley}, \citenamefont {Gu{\'{e}}nault}, \citenamefont {Gunnarsson},
		\citenamefont {Haley}, \citenamefont {Holt}, \citenamefont {Jones},
		\citenamefont {Pashkin}, \citenamefont {Penttil\"{a}}, \citenamefont
		{Prance}, \citenamefont {Prunnila},\ and\ \citenamefont
		{Roschier}}]{SciRepBradley2017}%
	\BibitemOpen
	\bibfield  {author} {\bibinfo {author} {\bibfnamefont {D.~I.}\ \bibnamefont
			{Bradley}}, \bibinfo {author} {\bibfnamefont {A.~M.}\ \bibnamefont
			{Gu{\'{e}}nault}}, \bibinfo {author} {\bibfnamefont {D.}~\bibnamefont
			{Gunnarsson}}, \bibinfo {author} {\bibfnamefont {R.~P.}\ \bibnamefont
			{Haley}}, \bibinfo {author} {\bibfnamefont {S.}~\bibnamefont {Holt}},
		\bibinfo {author} {\bibfnamefont {A.~T.}\ \bibnamefont {Jones}}, \bibinfo
		{author} {\bibfnamefont {Y.~A.}\ \bibnamefont {Pashkin}}, \bibinfo {author}
		{\bibfnamefont {J.}~\bibnamefont {Penttil\"{a}}}, \bibinfo {author}
		{\bibfnamefont {J.~R.}\ \bibnamefont {Prance}}, \bibinfo {author}
		{\bibfnamefont {M.}~\bibnamefont {Prunnila}},\ and\ \bibinfo {author}
		{\bibfnamefont {L.}~\bibnamefont {Roschier}},\ }\bibfield  {title} {\bibinfo
		{title} {On-chip magnetic cooling of a nanoelectronic device},\ }\href
	{https://doi.org/10.1038/srep45566} {\bibfield  {journal} {\bibinfo
			{journal} {Sci. Rep.}\ }\textbf {\bibinfo {volume} {7}},\ \bibinfo {pages}
		{45566} (\bibinfo {year} {2017})}\BibitemShut {NoStop}%
	\bibitem [{\citenamefont {Ziabari}\ \emph {et~al.}(2016)\citenamefont
		{Ziabari}, \citenamefont {Zebarjadi}, \citenamefont {Vashaee},\ and\
		\citenamefont {Shakouri}}]{Ziabari2016}%
	\BibitemOpen
	\bibfield  {author} {\bibinfo {author} {\bibfnamefont {A.}~\bibnamefont
			{Ziabari}}, \bibinfo {author} {\bibfnamefont {M.}~\bibnamefont {Zebarjadi}},
		\bibinfo {author} {\bibfnamefont {D.}~\bibnamefont {Vashaee}},\ and\ \bibinfo
		{author} {\bibfnamefont {A.}~\bibnamefont {Shakouri}},\ }\bibfield  {title}
	{\bibinfo {title} {Nanoscale solid-state cooling: a review},\ }\href
	{https://doi.org/10.1088/0034-4885/79/9/095901} {\bibfield  {journal}
		{\bibinfo  {journal} {Rep. Prog. Phys.}\ }\textbf {\bibinfo {volume} {79}},\
		\bibinfo {pages} {095901} (\bibinfo {year} {2016})}\BibitemShut {NoStop}%
	\bibitem [{\citenamefont {Shakouri}(2006)}]{Shakouri2006}%
	\BibitemOpen
	\bibfield  {author} {\bibinfo {author} {\bibfnamefont {A.}~\bibnamefont
			{Shakouri}},\ }\bibfield  {title} {\bibinfo {title} {Nanoscale thermal
			transport and microrefrigerators on a chip},\ }\href
	{https://doi.org/10.1109/JPROC.2006.879787} {\bibfield  {journal} {\bibinfo
			{journal} {Proceedings of the IEEE}\ }\textbf {\bibinfo {volume} {94}},\
		\bibinfo {pages} {1613} (\bibinfo {year} {2006})}\BibitemShut {NoStop}%
	\bibitem [{\citenamefont {Prance}\ \emph {et~al.}(2009)\citenamefont {Prance},
		\citenamefont {Smith}, \citenamefont {Griffiths}, \citenamefont {Chorley},
		\citenamefont {Anderson}, \citenamefont {Jones}, \citenamefont {Farrer},\
		and\ \citenamefont {Ritchie}}]{PrancePRL}%
	\BibitemOpen
	\bibfield  {author} {\bibinfo {author} {\bibfnamefont {J.~R.}\ \bibnamefont
			{Prance}}, \bibinfo {author} {\bibfnamefont {C.~G.}\ \bibnamefont {Smith}},
		\bibinfo {author} {\bibfnamefont {J.~P.}\ \bibnamefont {Griffiths}}, \bibinfo
		{author} {\bibfnamefont {S.~J.}\ \bibnamefont {Chorley}}, \bibinfo {author}
		{\bibfnamefont {D.}~\bibnamefont {Anderson}}, \bibinfo {author}
		{\bibfnamefont {G.~A.~C.}\ \bibnamefont {Jones}}, \bibinfo {author}
		{\bibfnamefont {I.}~\bibnamefont {Farrer}},\ and\ \bibinfo {author}
		{\bibfnamefont {D.~A.}\ \bibnamefont {Ritchie}},\ }\bibfield  {title}
	{\bibinfo {title} {Electronic refrigeration of a two-dimensional electron
			gas},\ }\href {https://doi.org/10.1103/PhysRevLett.102.146602} {\bibfield
		{journal} {\bibinfo  {journal} {Phys. Rev. Lett.}\ }\textbf {\bibinfo
			{volume} {102}},\ \bibinfo {pages} {146602} (\bibinfo {year}
		{2009})}\BibitemShut {NoStop}%
	\bibitem [{\citenamefont {Bell}(2008)}]{Bell2008}%
	\BibitemOpen
	\bibfield  {author} {\bibinfo {author} {\bibfnamefont {L.~E.}\ \bibnamefont
			{Bell}},\ }\bibfield  {title} {\bibinfo {title} {Cooling, heating, generating
			power, and recovering waste heat with thermoelectric systems},\ }\href
	{https://doi.org/10.1126/science.1158899} {\bibfield  {journal} {\bibinfo
			{journal} {Science}\ }\textbf {\bibinfo {volume} {321}},\ \bibinfo {pages}
		{1457} (\bibinfo {year} {2008})}\BibitemShut {NoStop}%
	\bibitem [{\citenamefont {Sothmann}\ \emph {et~al.}(2014)\citenamefont
		{Sothmann}, \citenamefont {S{\'{a}}nchez},\ and\ \citenamefont
		{Jordan}}]{Sothmann_2014}%
	\BibitemOpen
	\bibfield  {author} {\bibinfo {author} {\bibfnamefont {B.}~\bibnamefont
			{Sothmann}}, \bibinfo {author} {\bibfnamefont {R.}~\bibnamefont
			{S{\'{a}}nchez}},\ and\ \bibinfo {author} {\bibfnamefont {A.~N.}\
			\bibnamefont {Jordan}},\ }\bibfield  {title} {\bibinfo {title}
		{Thermoelectric energy harvesting with quantum dots},\ }\href
	{https://doi.org/10.1088/0957-4484/26/3/032001} {\bibfield  {journal}
		{\bibinfo  {journal} {Nanotechnology}\ }\textbf {\bibinfo {volume} {26}},\
		\bibinfo {pages} {032001} (\bibinfo {year} {2014})}\BibitemShut {NoStop}%
	\bibitem [{\citenamefont {Jaliel}\ \emph {et~al.}(2019)\citenamefont {Jaliel},
		\citenamefont {Puddy}, \citenamefont {S\'anchez}, \citenamefont {Jordan},
		\citenamefont {Sothmann}, \citenamefont {Farrer}, \citenamefont {Griffiths},
		\citenamefont {Ritchie},\ and\ \citenamefont {Smith}}]{QDHarvesterPRL}%
	\BibitemOpen
	\bibfield  {author} {\bibinfo {author} {\bibfnamefont {G.}~\bibnamefont
			{Jaliel}}, \bibinfo {author} {\bibfnamefont {R.~K.}\ \bibnamefont {Puddy}},
		\bibinfo {author} {\bibfnamefont {R.}~\bibnamefont {S\'anchez}}, \bibinfo
		{author} {\bibfnamefont {A.~N.}\ \bibnamefont {Jordan}}, \bibinfo {author}
		{\bibfnamefont {B.}~\bibnamefont {Sothmann}}, \bibinfo {author}
		{\bibfnamefont {I.}~\bibnamefont {Farrer}}, \bibinfo {author} {\bibfnamefont
			{J.~P.}\ \bibnamefont {Griffiths}}, \bibinfo {author} {\bibfnamefont {D.~A.}\
			\bibnamefont {Ritchie}},\ and\ \bibinfo {author} {\bibfnamefont {C.~G.}\
			\bibnamefont {Smith}},\ }\bibfield  {title} {\bibinfo {title} {Experimental
			realization of a quantum dot energy harvester},\ }\href
	{https://doi.org/10.1103/PhysRevLett.123.117701} {\bibfield  {journal}
		{\bibinfo  {journal} {Phys. Rev. Lett.}\ }\textbf {\bibinfo {volume} {123}},\
		\bibinfo {pages} {117701} (\bibinfo {year} {2019})}\BibitemShut {NoStop}%
	\bibitem [{\citenamefont {Thierschmann}\ \emph {et~al.}(2015)\citenamefont
		{Thierschmann}, \citenamefont {S{\'{a}}nchez}, \citenamefont {Sothmann},
		\citenamefont {Arnold}, \citenamefont {Heyn}, \citenamefont {Hansen},
		\citenamefont {Buhmann},\ and\ \citenamefont {Molenkamp}}]{Thierschmann2015}%
	\BibitemOpen
	\bibfield  {author} {\bibinfo {author} {\bibfnamefont {H.}~\bibnamefont
			{Thierschmann}}, \bibinfo {author} {\bibfnamefont {R.}~\bibnamefont
			{S{\'{a}}nchez}}, \bibinfo {author} {\bibfnamefont {B.}~\bibnamefont
			{Sothmann}}, \bibinfo {author} {\bibfnamefont {F.}~\bibnamefont {Arnold}},
		\bibinfo {author} {\bibfnamefont {C.}~\bibnamefont {Heyn}}, \bibinfo {author}
		{\bibfnamefont {W.}~\bibnamefont {Hansen}}, \bibinfo {author} {\bibfnamefont
			{H.}~\bibnamefont {Buhmann}},\ and\ \bibinfo {author} {\bibfnamefont {L.~W.}\
			\bibnamefont {Molenkamp}},\ }\bibfield  {title} {\bibinfo {title}
		{Three-terminal energy harvester with coupled quantum dots},\ }\href
	{https://doi.org/10.1038/nnano.2015.176} {\bibfield  {journal} {\bibinfo
			{journal} {Nat. Nanotech.}\ }\textbf {\bibinfo {volume} {10}},\ \bibinfo
		{pages} {854} (\bibinfo {year} {2015})}\BibitemShut {NoStop}%
	\bibitem [{\citenamefont {Roche}\ \emph {et~al.}(2015)\citenamefont {Roche},
		\citenamefont {Roulleau}, \citenamefont {Jullien}, \citenamefont {Jompol},
		\citenamefont {Farrer}, \citenamefont {Ritchie},\ and\ \citenamefont
		{Glattli}}]{Roche2015}%
	\BibitemOpen
	\bibfield  {author} {\bibinfo {author} {\bibfnamefont {B.}~\bibnamefont
			{Roche}}, \bibinfo {author} {\bibfnamefont {P.}~\bibnamefont {Roulleau}},
		\bibinfo {author} {\bibfnamefont {T.}~\bibnamefont {Jullien}}, \bibinfo
		{author} {\bibfnamefont {Y.}~\bibnamefont {Jompol}}, \bibinfo {author}
		{\bibfnamefont {I.}~\bibnamefont {Farrer}}, \bibinfo {author} {\bibfnamefont
			{D.}~\bibnamefont {Ritchie}},\ and\ \bibinfo {author} {\bibfnamefont
			{D.}~\bibnamefont {Glattli}},\ }\bibfield  {title} {\bibinfo {title}
		{Harvesting dissipated energy with a mesoscopic ratchet},\ }\href
	{https://doi.org/10.1038/ncomms7738} {\bibfield  {journal} {\bibinfo
			{journal} {Nat. Commun.}\ }\textbf {\bibinfo {volume} {6}},\ \bibinfo {pages}
		{6738} (\bibinfo {year} {2015})}\BibitemShut {NoStop}%
	\bibitem [{\citenamefont {Hartmann}\ \emph {et~al.}(2015)\citenamefont
		{Hartmann}, \citenamefont {Pfeffer}, \citenamefont {H\"ofling}, \citenamefont
		{Kamp},\ and\ \citenamefont {Worschech}}]{HartmannPRL2015}%
	\BibitemOpen
	\bibfield  {author} {\bibinfo {author} {\bibfnamefont {F.}~\bibnamefont
			{Hartmann}}, \bibinfo {author} {\bibfnamefont {P.}~\bibnamefont {Pfeffer}},
		\bibinfo {author} {\bibfnamefont {S.}~\bibnamefont {H\"ofling}}, \bibinfo
		{author} {\bibfnamefont {M.}~\bibnamefont {Kamp}},\ and\ \bibinfo {author}
		{\bibfnamefont {L.}~\bibnamefont {Worschech}},\ }\bibfield  {title} {\bibinfo
		{title} {Voltage fluctuation to current converter with coulomb-coupled
			quantum dots},\ }\href {https://doi.org/10.1103/PhysRevLett.114.146805}
	{\bibfield  {journal} {\bibinfo  {journal} {Phys. Rev. Lett.}\ }\textbf
		{\bibinfo {volume} {114}},\ \bibinfo {pages} {146805} (\bibinfo {year}
		{2015})}\BibitemShut {NoStop}%
	\bibitem [{\citenamefont {Goldsmid}(2016)}]{Goldsmid2016}%
	\BibitemOpen
	\bibfield  {author} {\bibinfo {author} {\bibfnamefont {H.~J.}\ \bibnamefont
			{Goldsmid}},\ }\href {https://doi.org/10.1007/978-3-662-49256-7} {\emph
		{\bibinfo {title} {Introduction to Thermoelectricity}}}\ (\bibinfo
	{publisher} {Springer Berlin Heidelberg},\ \bibinfo {year}
	{2016})\BibitemShut {NoStop}%
	\bibitem [{\citenamefont {S\'anchez}\ and\ \citenamefont
		{L\'opez}(2016)}]{SanchezNonlinearReview}%
	\BibitemOpen
	\bibfield  {author} {\bibinfo {author} {\bibfnamefont {D.}~\bibnamefont
			{S\'anchez}}\ and\ \bibinfo {author} {\bibfnamefont {R.}~\bibnamefont
			{L\'opez}},\ }\bibfield  {title} {\bibinfo {title} {Nonlinear phenomena in
			quantum thermoelectrics and heat},\ }\href
	{https://doi.org/https://doi.org/10.1016/j.crhy.2016.08.005} {\bibfield
		{journal} {\bibinfo  {journal} {C. R. Physique}\ }\textbf {\bibinfo {volume}
			{17}},\ \bibinfo {pages} {1060 } (\bibinfo {year} {2016})}\BibitemShut
	{NoStop}%
	\bibitem [{\citenamefont {Reddy}\ \emph {et~al.}(2007)\citenamefont {Reddy},
		\citenamefont {Jang}, \citenamefont {Segalman},\ and\ \citenamefont
		{Majumdar}}]{Reddy2007}%
	\BibitemOpen
	\bibfield  {author} {\bibinfo {author} {\bibfnamefont {P.}~\bibnamefont
			{Reddy}}, \bibinfo {author} {\bibfnamefont {S.-Y.}\ \bibnamefont {Jang}},
		\bibinfo {author} {\bibfnamefont {R.~A.}\ \bibnamefont {Segalman}},\ and\
		\bibinfo {author} {\bibfnamefont {A.}~\bibnamefont {Majumdar}},\ }\bibfield
	{title} {\bibinfo {title} {Thermoelectricity in molecular junctions},\ }\href
	{https://doi.org/10.1126/science.1137149} {\bibfield  {journal} {\bibinfo
			{journal} {Science}\ }\textbf {\bibinfo {volume} {315}},\ \bibinfo {pages}
		{1568} (\bibinfo {year} {2007})}\BibitemShut {NoStop}%
	\bibitem [{\citenamefont {Brantut}\ \emph {et~al.}(2013)\citenamefont
		{Brantut}, \citenamefont {Grenier}, \citenamefont {Meineke}, \citenamefont
		{Stadler}, \citenamefont {Krinner}, \citenamefont {Kollath}, \citenamefont
		{Esslinger},\ and\ \citenamefont {Georges}}]{Brantut2013}%
	\BibitemOpen
	\bibfield  {author} {\bibinfo {author} {\bibfnamefont {J.-P.}\ \bibnamefont
			{Brantut}}, \bibinfo {author} {\bibfnamefont {C.}~\bibnamefont {Grenier}},
		\bibinfo {author} {\bibfnamefont {J.}~\bibnamefont {Meineke}}, \bibinfo
		{author} {\bibfnamefont {D.}~\bibnamefont {Stadler}}, \bibinfo {author}
		{\bibfnamefont {S.}~\bibnamefont {Krinner}}, \bibinfo {author} {\bibfnamefont
			{C.}~\bibnamefont {Kollath}}, \bibinfo {author} {\bibfnamefont
			{T.}~\bibnamefont {Esslinger}},\ and\ \bibinfo {author} {\bibfnamefont
			{A.}~\bibnamefont {Georges}},\ }\bibfield  {title} {\bibinfo {title} {A
			thermoelectric heat engine with ultracold atoms},\ }\href
	{https://doi.org/10.1126/science.1242308} {\bibfield  {journal} {\bibinfo
			{journal} {Science}\ }\textbf {\bibinfo {volume} {342}},\ \bibinfo {pages}
		{713} (\bibinfo {year} {2013})}\BibitemShut {NoStop}%
	\bibitem [{\citenamefont {Josefsson}\ \emph {et~al.}(2018)\citenamefont
		{Josefsson}, \citenamefont {Svilans}, \citenamefont {Burke}, \citenamefont
		{Hoffmann}, \citenamefont {Fahlvik}, \citenamefont {Thelander}, \citenamefont
		{Leijnse},\ and\ \citenamefont {Linke}}]{Josefsson2018}%
	\BibitemOpen
	\bibfield  {author} {\bibinfo {author} {\bibfnamefont {M.}~\bibnamefont
			{Josefsson}}, \bibinfo {author} {\bibfnamefont {A.}~\bibnamefont {Svilans}},
		\bibinfo {author} {\bibfnamefont {A.~M.}\ \bibnamefont {Burke}}, \bibinfo
		{author} {\bibfnamefont {E.~A.}\ \bibnamefont {Hoffmann}}, \bibinfo {author}
		{\bibfnamefont {S.}~\bibnamefont {Fahlvik}}, \bibinfo {author} {\bibfnamefont
			{C.}~\bibnamefont {Thelander}}, \bibinfo {author} {\bibfnamefont
			{M.}~\bibnamefont {Leijnse}},\ and\ \bibinfo {author} {\bibfnamefont
			{H.}~\bibnamefont {Linke}},\ }\bibfield  {title} {\bibinfo {title} {A
			quantum-dot heat engine operating close to the thermodynamic efficiency
			limits},\ }\href {https://doi.org/10.1038/s41565-018-0200-5} {\bibfield
		{journal} {\bibinfo  {journal} {Nat. Nanotech.}\ }\textbf {\bibinfo {volume}
			{13}},\ \bibinfo {pages} {920} (\bibinfo {year} {2018})}\BibitemShut
	{NoStop}%
	\bibitem [{\citenamefont {Kurizki}\ \emph {et~al.}(2015)\citenamefont
		{Kurizki}, \citenamefont {Bertet}, \citenamefont {Kubo}, \citenamefont
		{M{\o}lmer}, \citenamefont {Petrosyan}, \citenamefont {Rabl},\ and\
		\citenamefont {Schmiedmayer}}]{Kurizki3866}%
	\BibitemOpen
	\bibfield  {author} {\bibinfo {author} {\bibfnamefont {G.}~\bibnamefont
			{Kurizki}}, \bibinfo {author} {\bibfnamefont {P.}~\bibnamefont {Bertet}},
		\bibinfo {author} {\bibfnamefont {Y.}~\bibnamefont {Kubo}}, \bibinfo {author}
		{\bibfnamefont {K.}~\bibnamefont {M{\o}lmer}}, \bibinfo {author}
		{\bibfnamefont {D.}~\bibnamefont {Petrosyan}}, \bibinfo {author}
		{\bibfnamefont {P.}~\bibnamefont {Rabl}},\ and\ \bibinfo {author}
		{\bibfnamefont {J.}~\bibnamefont {Schmiedmayer}},\ }\bibfield  {title}
	{\bibinfo {title} {Quantum technologies with hybrid systems},\ }\href
	{https://doi.org/10.1073/pnas.1419326112} {\bibfield  {journal} {\bibinfo
			{journal} {PNAS}\ }\textbf {\bibinfo {volume} {112}},\ \bibinfo {pages}
		{3866} (\bibinfo {year} {2015})}\BibitemShut {NoStop}%
	\bibitem [{\citenamefont {Pekola}(2015)}]{PekolaReview2015}%
	\BibitemOpen
	\bibfield  {author} {\bibinfo {author} {\bibfnamefont {J.~P.}\ \bibnamefont
			{Pekola}},\ }\bibfield  {title} {\bibinfo {title} {Towards quantum
			thermodynamics in electronic~circuits},\ }\href
	{https://doi.org/10.1038/nphys3169} {\bibfield  {journal} {\bibinfo
			{journal} {Nat. Phys.}\ }\textbf {\bibinfo {volume} {11}},\ \bibinfo {pages}
		{118} (\bibinfo {year} {2015})}\BibitemShut {NoStop}%
	\bibitem [{\citenamefont {Wendin}(2017)}]{Wendin2017}%
	\BibitemOpen
	\bibfield  {author} {\bibinfo {author} {\bibfnamefont {G.}~\bibnamefont
			{Wendin}},\ }\bibfield  {title} {\bibinfo {title} {Quantum information
			processing with superconducting circuits: a review},\ }\href
	{https://doi.org/10.1088/1361-6633/aa7e1a} {\bibfield  {journal} {\bibinfo
			{journal} {Rep. Prog. Phys.}\ }\textbf {\bibinfo {volume} {80}},\ \bibinfo
		{pages} {106001} (\bibinfo {year} {2017})}\BibitemShut {NoStop}%
	\bibitem [{\citenamefont {Krantz}\ \emph {et~al.}(2019)\citenamefont {Krantz},
		\citenamefont {Kjaergaard}, \citenamefont {Yan}, \citenamefont {Orlando},
		\citenamefont {Gustavsson},\ and\ \citenamefont {Oliver}}]{Krantz2019}%
	\BibitemOpen
	\bibfield  {author} {\bibinfo {author} {\bibfnamefont {P.}~\bibnamefont
			{Krantz}}, \bibinfo {author} {\bibfnamefont {M.}~\bibnamefont {Kjaergaard}},
		\bibinfo {author} {\bibfnamefont {F.}~\bibnamefont {Yan}}, \bibinfo {author}
		{\bibfnamefont {T.~P.}\ \bibnamefont {Orlando}}, \bibinfo {author}
		{\bibfnamefont {S.}~\bibnamefont {Gustavsson}},\ and\ \bibinfo {author}
		{\bibfnamefont {W.~D.}\ \bibnamefont {Oliver}},\ }\bibfield  {title}
	{\bibinfo {title} {A quantum engineer's guide to superconducting qubits},\
	}\href {https://doi.org/10.1063/1.5089550} {\bibfield  {journal} {\bibinfo
			{journal} {Appl. Phys. Rev.}\ }\textbf {\bibinfo {volume} {6}},\ \bibinfo
		{pages} {021318} (\bibinfo {year} {2019})}\BibitemShut {NoStop}%
	\bibitem [{\citenamefont {Fornieri}\ and\ \citenamefont
		{Giazotto}(2017)}]{FornieriReview}%
	\BibitemOpen
	\bibfield  {author} {\bibinfo {author} {\bibfnamefont {A.}~\bibnamefont
			{Fornieri}}\ and\ \bibinfo {author} {\bibfnamefont {F.}~\bibnamefont
			{Giazotto}},\ }\bibfield  {title} {\bibinfo {title} {Towards phase-coherent
			caloritronics in superconducting circuits},\ }\href@noop {} {\bibfield
		{journal} {\bibinfo  {journal} {Nat. Nanotechnol.}\ }\textbf {\bibinfo
			{volume} {12}},\ \bibinfo {pages} {944} (\bibinfo {year} {2017})}\BibitemShut
	{NoStop}%
	\bibitem [{\citenamefont {Hwang}\ and\ \citenamefont
		{Sothmann}(2020)}]{Hwang2020}%
	\BibitemOpen
	\bibfield  {author} {\bibinfo {author} {\bibfnamefont {S.-Y.}\ \bibnamefont
			{Hwang}}\ and\ \bibinfo {author} {\bibfnamefont {B.}~\bibnamefont
			{Sothmann}},\ }\bibfield  {title} {\bibinfo {title} {Phase-coherent
			caloritronics with ordinary and topological josephson junctions},\ }\href
	{https://doi.org/10.1140/epjst/e2019-900094-y} {\bibfield  {journal}
		{\bibinfo  {journal} {Eur. Phys. J. Special Topics}\ }\textbf {\bibinfo
			{volume} {229}},\ \bibinfo {pages} {683} (\bibinfo {year}
		{2020})}\BibitemShut {NoStop}%
	\bibitem [{\citenamefont {Machon}\ \emph {et~al.}(2013)\citenamefont {Machon},
		\citenamefont {Eschrig},\ and\ \citenamefont {Belzig}}]{BelzigTEPRL}%
	\BibitemOpen
	\bibfield  {author} {\bibinfo {author} {\bibfnamefont {P.}~\bibnamefont
			{Machon}}, \bibinfo {author} {\bibfnamefont {M.}~\bibnamefont {Eschrig}},\
		and\ \bibinfo {author} {\bibfnamefont {W.}~\bibnamefont {Belzig}},\
	}\bibfield  {title} {\bibinfo {title} {Nonlocal thermoelectric effects and
			nonlocal onsager relations in a three-terminal proximity-coupled
			superconductor-ferromagnet device},\ }\href
	{https://doi.org/10.1103/PhysRevLett.110.047002} {\bibfield  {journal}
		{\bibinfo  {journal} {Phys. Rev. Lett.}\ }\textbf {\bibinfo {volume} {110}},\
		\bibinfo {pages} {047002} (\bibinfo {year} {2013})}\BibitemShut {NoStop}%
	\bibitem [{\citenamefont {Ozaeta}\ \emph {et~al.}(2014)\citenamefont {Ozaeta},
		\citenamefont {Virtanen}, \citenamefont {Bergeret},\ and\ \citenamefont
		{Heikkil{\"a}}}]{Ozaeta2014}%
	\BibitemOpen
	\bibfield  {author} {\bibinfo {author} {\bibfnamefont {A.}~\bibnamefont
			{Ozaeta}}, \bibinfo {author} {\bibfnamefont {P.}~\bibnamefont {Virtanen}},
		\bibinfo {author} {\bibfnamefont {F.~S.}\ \bibnamefont {Bergeret}},\ and\
		\bibinfo {author} {\bibfnamefont {T.~T.}\ \bibnamefont {Heikkil{\"a}}},\
	}\bibfield  {title} {\bibinfo {title} {{Predicted very large thermoelectric
				effect in ferromagnet-superconductor junctions in the presence of a
				spin-splitting magnetic field}},\ }\href
	{https://doi.org/10.1103/PhysRevLett.112.057001} {\bibfield  {journal}
		{\bibinfo  {journal} {Phys. Rev. Lett.}\ }\textbf {\bibinfo {volume} {112}},\
		\bibinfo {pages} {057001} (\bibinfo {year} {2014})}\BibitemShut {NoStop}%
	\bibitem [{\citenamefont {Kolenda}\ \emph {et~al.}(2017)\citenamefont
		{Kolenda}, \citenamefont {S\"urgers}, \citenamefont {Fischer},\ and\
		\citenamefont {Beckmann}}]{Kolenda2017}%
	\BibitemOpen
	\bibfield  {author} {\bibinfo {author} {\bibfnamefont {S.}~\bibnamefont
			{Kolenda}}, \bibinfo {author} {\bibfnamefont {C.}~\bibnamefont {S\"urgers}},
		\bibinfo {author} {\bibfnamefont {G.}~\bibnamefont {Fischer}},\ and\ \bibinfo
		{author} {\bibfnamefont {D.}~\bibnamefont {Beckmann}},\ }\bibfield  {title}
	{\bibinfo {title} {Thermoelectric effects in superconductor-ferromagnet
			tunnel junctions on europium sulfide},\ }\href
	{https://doi.org/10.1103/PhysRevB.95.224505} {\bibfield  {journal} {\bibinfo
			{journal} {Phys. Rev. B}\ }\textbf {\bibinfo {volume} {95}},\ \bibinfo
		{pages} {224505} (\bibinfo {year} {2017})}\BibitemShut {NoStop}%
	\bibitem [{\citenamefont {Beckmann}(2016)}]{Beckmann2016}%
	\BibitemOpen
	\bibfield  {author} {\bibinfo {author} {\bibfnamefont {D.}~\bibnamefont
			{Beckmann}},\ }\bibfield  {title} {\bibinfo {title} {{Spin manipulation in
				nanoscale superconductors}},\ }\href
	{https://doi.org/10.1088/0953-8984/28/16/163001} {\bibfield  {journal}
		{\bibinfo  {journal} {J. Phys. Condens. Matter}\ }\textbf {\bibinfo {volume}
			{28}},\ \bibinfo {pages} {163001} (\bibinfo {year} {2016})}\BibitemShut
	{NoStop}%
	\bibitem [{\citenamefont {Hussein}\ \emph {et~al.}(2019)\citenamefont
		{Hussein}, \citenamefont {Governale}, \citenamefont {Kohler}, \citenamefont
		{Belzig}, \citenamefont {Giazotto},\ and\ \citenamefont
		{Braggio}}]{BraggioNonlocalTE}%
	\BibitemOpen
	\bibfield  {author} {\bibinfo {author} {\bibfnamefont {R.}~\bibnamefont
			{Hussein}}, \bibinfo {author} {\bibfnamefont {M.}~\bibnamefont {Governale}},
		\bibinfo {author} {\bibfnamefont {S.}~\bibnamefont {Kohler}}, \bibinfo
		{author} {\bibfnamefont {W.}~\bibnamefont {Belzig}}, \bibinfo {author}
		{\bibfnamefont {F.}~\bibnamefont {Giazotto}},\ and\ \bibinfo {author}
		{\bibfnamefont {A.}~\bibnamefont {Braggio}},\ }\bibfield  {title} {\bibinfo
		{title} {Nonlocal thermoelectricity in a cooper-pair splitter},\ }\href
	{https://doi.org/10.1103/PhysRevB.99.075429} {\bibfield  {journal} {\bibinfo
			{journal} {Phys. Rev. B}\ }\textbf {\bibinfo {volume} {99}},\ \bibinfo
		{pages} {075429} (\bibinfo {year} {2019})}\BibitemShut {NoStop}%
	\bibitem [{\citenamefont {Pershoguba}\ and\ \citenamefont
		{Glazman}(2019)}]{GlazmanPRB99}%
	\BibitemOpen
	\bibfield  {author} {\bibinfo {author} {\bibfnamefont {S.~S.}\ \bibnamefont
			{Pershoguba}}\ and\ \bibinfo {author} {\bibfnamefont {L.~I.}\ \bibnamefont
			{Glazman}},\ }\bibfield  {title} {\bibinfo {title} {Thermopower and thermal
			conductance of a superconducting quantum point contact},\ }\href
	{https://doi.org/10.1103/PhysRevB.99.134514} {\bibfield  {journal} {\bibinfo
			{journal} {Phys. Rev. B}\ }\textbf {\bibinfo {volume} {99}},\ \bibinfo
		{pages} {134514} (\bibinfo {year} {2019})}\BibitemShut {NoStop}%
	\bibitem [{\citenamefont {Kirsanov}\ \emph {et~al.}(2019)\citenamefont
		{Kirsanov}, \citenamefont {Tan}, \citenamefont {Golubev}, \citenamefont
		{Hakonen},\ and\ \citenamefont {Lesovik}}]{LesovikPRB99}%
	\BibitemOpen
	\bibfield  {author} {\bibinfo {author} {\bibfnamefont {N.~S.}\ \bibnamefont
			{Kirsanov}}, \bibinfo {author} {\bibfnamefont {Z.~B.}\ \bibnamefont {Tan}},
		\bibinfo {author} {\bibfnamefont {D.~S.}\ \bibnamefont {Golubev}}, \bibinfo
		{author} {\bibfnamefont {P.~J.}\ \bibnamefont {Hakonen}},\ and\ \bibinfo
		{author} {\bibfnamefont {G.~B.}\ \bibnamefont {Lesovik}},\ }\bibfield
	{title} {\bibinfo {title} {Heat switch and thermoelectric effects based on
			cooper-pair splitting and elastic cotunneling},\ }\href
	{https://doi.org/10.1103/PhysRevB.99.115127} {\bibfield  {journal} {\bibinfo
			{journal} {Phys. Rev. B}\ }\textbf {\bibinfo {volume} {99}},\ \bibinfo
		{pages} {115127} (\bibinfo {year} {2019})}\BibitemShut {NoStop}%
	\bibitem [{\citenamefont {Blasi}\ \emph {et~al.}()\citenamefont {Blasi},
		\citenamefont {Taddei}, \citenamefont {Arrachea}, \citenamefont {Carrega},\
		and\ \citenamefont {Braggio}}]{Blasi}%
	\BibitemOpen
	\bibfield  {author} {\bibinfo {author} {\bibfnamefont {G.}~\bibnamefont
			{Blasi}}, \bibinfo {author} {\bibfnamefont {F.}~\bibnamefont {Taddei}},
		\bibinfo {author} {\bibfnamefont {L.}~\bibnamefont {Arrachea}}, \bibinfo
		{author} {\bibfnamefont {M.}~\bibnamefont {Carrega}},\ and\ \bibinfo {author}
		{\bibfnamefont {A.}~\bibnamefont {Braggio}},\ }\bibfield  {title} {\bibinfo
		{title} {Nonlocal thermoelectric signature of helical edge states},\
	}\href@noop {} {\bibfield  {journal} {\bibinfo  {journal} {arXiv:preprint}\
	}}\Eprint {https://arxiv.org/abs/1911.04367} {arXiv:1911.04367
		[cond-mat.mes-hall]} \BibitemShut {NoStop}%
	\bibitem [{\citenamefont {Giazotto}\ \emph {et~al.}(2015)\citenamefont
		{Giazotto}, \citenamefont {Solinas}, \citenamefont {Braggio},\ and\
		\citenamefont {Bergeret}}]{GiazottoThermometerNFIS}%
	\BibitemOpen
	\bibfield  {author} {\bibinfo {author} {\bibfnamefont {F.}~\bibnamefont
			{Giazotto}}, \bibinfo {author} {\bibfnamefont {P.}~\bibnamefont {Solinas}},
		\bibinfo {author} {\bibfnamefont {A.}~\bibnamefont {Braggio}},\ and\ \bibinfo
		{author} {\bibfnamefont {F.~S.}\ \bibnamefont {Bergeret}},\ }\bibfield
	{title} {\bibinfo {title} {Ferromagnetic-insulator-based superconducting
			junctions as sensitive electron thermometers},\ }\href
	{https://doi.org/10.1103/PhysRevApplied.4.044016} {\bibfield  {journal}
		{\bibinfo  {journal} {Phys. Rev. Applied}\ }\textbf {\bibinfo {volume} {4}},\
		\bibinfo {pages} {044016} (\bibinfo {year} {2015})}\BibitemShut {NoStop}%
	\bibitem [{\citenamefont {Marchegiani}\ \emph {et~al.}(2016)\citenamefont
		{Marchegiani}, \citenamefont {Virtanen}, \citenamefont {Giazotto},\ and\
		\citenamefont {Campisi}}]{MarchegianiEngine}%
	\BibitemOpen
	\bibfield  {author} {\bibinfo {author} {\bibfnamefont {G.}~\bibnamefont
			{Marchegiani}}, \bibinfo {author} {\bibfnamefont {P.}~\bibnamefont
			{Virtanen}}, \bibinfo {author} {\bibfnamefont {F.}~\bibnamefont {Giazotto}},\
		and\ \bibinfo {author} {\bibfnamefont {M.}~\bibnamefont {Campisi}},\
	}\bibfield  {title} {\bibinfo {title} {Self-oscillating josephson quantum
			heat engine},\ }\href {https://doi.org/10.1103/PhysRevApplied.6.054014}
	{\bibfield  {journal} {\bibinfo  {journal} {Phys. Rev. Applied}\ }\textbf
		{\bibinfo {volume} {6}},\ \bibinfo {pages} {054014} (\bibinfo {year}
		{2016})}\BibitemShut {NoStop}%
	\bibitem [{\citenamefont {Marchegiani}\ \emph {et~al.}(2018)\citenamefont
		{Marchegiani}, \citenamefont {Virtanen},\ and\ \citenamefont
		{Giazotto}}]{MarchegianiCooler}%
	\BibitemOpen
	\bibfield  {author} {\bibinfo {author} {\bibfnamefont {G.}~\bibnamefont
			{Marchegiani}}, \bibinfo {author} {\bibfnamefont {P.}~\bibnamefont
			{Virtanen}},\ and\ \bibinfo {author} {\bibfnamefont {F.}~\bibnamefont
			{Giazotto}},\ }\bibfield  {title} {\bibinfo {title} {On-chip cooling by
			heating with superconducting tunnel junctions},\ }\href
	{https://doi.org/10.1209/0295-5075/124/48005} {\bibfield  {journal} {\bibinfo
			{journal} {Europhys. Lett.}\ }\textbf {\bibinfo {volume} {124}},\ \bibinfo
		{pages} {48005} (\bibinfo {year} {2018})}\BibitemShut {NoStop}%
	\bibitem [{\citenamefont {Heikkil\"a}\ \emph {et~al.}(2018)\citenamefont
		{Heikkil\"a}, \citenamefont {Ojaj\"arvi}, \citenamefont {Maasilta},
		\citenamefont {Strambini}, \citenamefont {Giazotto},\ and\ \citenamefont
		{Bergeret}}]{GiazottoTEDetector}%
	\BibitemOpen
	\bibfield  {author} {\bibinfo {author} {\bibfnamefont {T.~T.}\ \bibnamefont
			{Heikkil\"a}}, \bibinfo {author} {\bibfnamefont {R.}~\bibnamefont
			{Ojaj\"arvi}}, \bibinfo {author} {\bibfnamefont {I.~J.}\ \bibnamefont
			{Maasilta}}, \bibinfo {author} {\bibfnamefont {E.}~\bibnamefont {Strambini}},
		\bibinfo {author} {\bibfnamefont {F.}~\bibnamefont {Giazotto}},\ and\
		\bibinfo {author} {\bibfnamefont {F.~S.}\ \bibnamefont {Bergeret}},\
	}\bibfield  {title} {\bibinfo {title} {Thermoelectric radiation detector
			based on superconductor-ferromagnet systems},\ }\href
	{https://doi.org/10.1103/PhysRevApplied.10.034053} {\bibfield  {journal}
		{\bibinfo  {journal} {Phys. Rev. Applied}\ }\textbf {\bibinfo {volume}
			{10}},\ \bibinfo {pages} {034053} (\bibinfo {year} {2018})}\BibitemShut
	{NoStop}%
	\bibitem [{\citenamefont {Strambini}\ \emph {et~al.}(2017)\citenamefont
		{Strambini}, \citenamefont {Golovach}, \citenamefont {De~Simoni},
		\citenamefont {Moodera}, \citenamefont {Bergeret},\ and\ \citenamefont
		{Giazotto}}]{Strambini2017}%
	\BibitemOpen
	\bibfield  {author} {\bibinfo {author} {\bibfnamefont {E.}~\bibnamefont
			{Strambini}}, \bibinfo {author} {\bibfnamefont {V.~N.}\ \bibnamefont
			{Golovach}}, \bibinfo {author} {\bibfnamefont {G.}~\bibnamefont {De~Simoni}},
		\bibinfo {author} {\bibfnamefont {J.~S.}\ \bibnamefont {Moodera}}, \bibinfo
		{author} {\bibfnamefont {F.~S.}\ \bibnamefont {Bergeret}},\ and\ \bibinfo
		{author} {\bibfnamefont {F.}~\bibnamefont {Giazotto}},\ }\bibfield  {title}
	{\bibinfo {title} {Revealing the magnetic proximity effect in eus/al bilayers
			through superconducting tunneling spectroscopy},\ }\href
	{https://doi.org/10.1103/PhysRevMaterials.1.054402} {\bibfield  {journal}
		{\bibinfo  {journal} {Phys. Rev. Materials}\ }\textbf {\bibinfo {volume}
			{1}},\ \bibinfo {pages} {054402} (\bibinfo {year} {2017})}\BibitemShut
	{NoStop}%
	\bibitem [{\citenamefont {Simoni}\ \emph {et~al.}(2018)\citenamefont {Simoni},
		\citenamefont {Strambini}, \citenamefont {Moodera}, \citenamefont
		{Bergeret},\ and\ \citenamefont {Giazotto}}]{DeSimoni2018}%
	\BibitemOpen
	\bibfield  {author} {\bibinfo {author} {\bibfnamefont {G.~D.}\ \bibnamefont
			{Simoni}}, \bibinfo {author} {\bibfnamefont {E.}~\bibnamefont {Strambini}},
		\bibinfo {author} {\bibfnamefont {J.~S.}\ \bibnamefont {Moodera}}, \bibinfo
		{author} {\bibfnamefont {F.~S.}\ \bibnamefont {Bergeret}},\ and\ \bibinfo
		{author} {\bibfnamefont {F.}~\bibnamefont {Giazotto}},\ }\bibfield  {title}
	{\bibinfo {title} {Toward the absolute spin-valve effect in superconducting
			tunnel junctions},\ }\href {https://doi.org/10.1021/acs.nanolett.8b02723}
	{\bibfield  {journal} {\bibinfo  {journal} {Nano Lett.}\ }\textbf {\bibinfo
			{volume} {18}},\ \bibinfo {pages} {6369} (\bibinfo {year}
		{2018})}\BibitemShut {NoStop}%
	\bibitem [{\citenamefont {Marchegiani}\ \emph {et~al.}(2020)\citenamefont
		{Marchegiani}, \citenamefont {Braggio},\ and\ \citenamefont
		{Giazotto}}]{MarchegianiNLTE}%
	\BibitemOpen
	\bibfield  {author} {\bibinfo {author} {\bibfnamefont {G.}~\bibnamefont
			{Marchegiani}}, \bibinfo {author} {\bibfnamefont {A.}~\bibnamefont
			{Braggio}},\ and\ \bibinfo {author} {\bibfnamefont {F.}~\bibnamefont
			{Giazotto}},\ }\href@noop {} {\bibinfo {title} {Nonlinear thermoelectricity
			with electron-hole symmetric systems}} (\bibinfo {year} {2020}),\ \bibinfo
	{note} {accepted by Phys. Rev. Lett., \textit{in press}}\BibitemShut
	{NoStop}%
	\bibitem [{\citenamefont {Tinkham}(2004)}]{Tinkham2004}%
	\BibitemOpen
	\bibfield  {author} {\bibinfo {author} {\bibfnamefont {M.}~\bibnamefont
			{Tinkham}},\ }\href@noop {} {\emph {\bibinfo {title} {{Introduction to
					superconductivity}}}}\ (\bibinfo  {publisher} {Dover Publications, Mineola,
		New York},\ \bibinfo {year} {2004})\BibitemShut {NoStop}%
	\bibitem [{\citenamefont {Barone}\ and\ \citenamefont
		{Patern{\`o}}(1982)}]{barone1982physics}%
	\BibitemOpen
	\bibfield  {author} {\bibinfo {author} {\bibfnamefont {A.}~\bibnamefont
			{Barone}}\ and\ \bibinfo {author} {\bibfnamefont {G.}~\bibnamefont
			{Patern{\`o}}},\ }\href@noop {} {\emph {\bibinfo {title} {Physics and
				applications of the Josephson effect}}}\ (\bibinfo  {publisher} {Wiley,New
		York},\ \bibinfo {year} {1982})\BibitemShut {NoStop}%
	\bibitem [{\citenamefont {Dynes}\ \emph {et~al.}(1978)\citenamefont {Dynes},
		\citenamefont {Narayanamurti},\ and\ \citenamefont {Garno}}]{Dynes1978}%
	\BibitemOpen
	\bibfield  {author} {\bibinfo {author} {\bibfnamefont {R.~C.}\ \bibnamefont
			{Dynes}}, \bibinfo {author} {\bibfnamefont {V.}~\bibnamefont
			{Narayanamurti}},\ and\ \bibinfo {author} {\bibfnamefont {J.~P.}\
			\bibnamefont {Garno}},\ }\bibfield  {title} {\bibinfo {title} {Direct
			measurement of quasiparticle-lifetime broadening in a strong-coupled
			superconductor},\ }\href {https://doi.org/10.1103/PhysRevLett.41.1509}
	{\bibfield  {journal} {\bibinfo  {journal} {Phys. Rev. Lett.}\ }\textbf
		{\bibinfo {volume} {41}},\ \bibinfo {pages} {1509} (\bibinfo {year}
		{1978})}\BibitemShut {NoStop}%
	\bibitem [{\citenamefont {Dynes}\ \emph {et~al.}(1984)\citenamefont {Dynes},
		\citenamefont {Garno}, \citenamefont {Hertel},\ and\ \citenamefont
		{Orlando}}]{Dynes1984}%
	\BibitemOpen
	\bibfield  {author} {\bibinfo {author} {\bibfnamefont {R.~C.}\ \bibnamefont
			{Dynes}}, \bibinfo {author} {\bibfnamefont {J.~P.}\ \bibnamefont {Garno}},
		\bibinfo {author} {\bibfnamefont {G.~B.}\ \bibnamefont {Hertel}},\ and\
		\bibinfo {author} {\bibfnamefont {T.~P.}\ \bibnamefont {Orlando}},\
	}\bibfield  {title} {\bibinfo {title} {{Tunneling Study of Superconductivity
				near the Metal-Insulator Transition}},\ }\href
	{https://doi.org/10.1103/PhysRevLett.53.2437} {\bibfield  {journal} {\bibinfo
			{journal} {Phys. Rev. Lett.}\ }\textbf {\bibinfo {volume} {53}},\ \bibinfo
		{pages} {2437} (\bibinfo {year} {1984})}\BibitemShut {NoStop}%
	\bibitem [{\citenamefont {Aronov}\ and\ \citenamefont
		{Spivak}(1975)}]{Aronov1975}%
	\BibitemOpen
	\bibfield  {author} {\bibinfo {author} {\bibfnamefont {A.~G.}\ \bibnamefont
			{Aronov}}\ and\ \bibinfo {author} {\bibfnamefont {B.~Z.}\ \bibnamefont
			{Spivak}},\ }\href@noop {} {\bibfield  {journal} {\bibinfo  {journal} {JETP
				Lett.}\ }\textbf {\bibinfo {volume} {22}},\ \bibinfo {pages} {101} (\bibinfo
		{year} {1975})}\BibitemShut {NoStop}%
	\bibitem [{\citenamefont {Gershenzon}\ and\ \citenamefont
		{Falei}(1986)}]{Gershenzon1986}%
	\BibitemOpen
	\bibfield  {author} {\bibinfo {author} {\bibfnamefont {M.~E.}\ \bibnamefont
			{Gershenzon}}\ and\ \bibinfo {author} {\bibfnamefont {M.~I.}\ \bibnamefont
			{Falei}},\ }\href@noop {} {\bibfield  {journal} {\bibinfo  {journal} {JETP
				Lett.}\ }\textbf {\bibinfo {volume} {44}},\ \bibinfo {pages} {682} (\bibinfo
		{year} {1986})}\BibitemShut {NoStop}%
	\bibitem [{\citenamefont {Gershenzon}\ and\ \citenamefont
		{Falei}(1988)}]{Gershenzon1988}%
	\BibitemOpen
	\bibfield  {author} {\bibinfo {author} {\bibfnamefont {M.~E.}\ \bibnamefont
			{Gershenzon}}\ and\ \bibinfo {author} {\bibfnamefont {M.~I.}\ \bibnamefont
			{Falei}},\ }\href@noop {} {\bibfield  {journal} {\bibinfo  {journal} {Sov.
				Phys. JETP}\ }\textbf {\bibinfo {volume} {67}},\ \bibinfo {pages} {389}
		(\bibinfo {year} {1988})}\BibitemShut {NoStop}%
	\bibitem [{\citenamefont {Gijsbertsen}\ and\ \citenamefont
		{Flokstra}(1996)}]{Gijsbertsen1996}%
	\BibitemOpen
	\bibfield  {author} {\bibinfo {author} {\bibfnamefont {J.~G.}\ \bibnamefont
			{Gijsbertsen}}\ and\ \bibinfo {author} {\bibfnamefont {J.}~\bibnamefont
			{Flokstra}},\ }\bibfield  {title} {\bibinfo {title} {Quasiparticle
			injection-detection experiments in niobium},\ }\href
	{https://doi.org/10.1063/1.363350} {\bibfield  {journal} {\bibinfo  {journal}
			{J. Appl. Phys.}\ }\textbf {\bibinfo {volume} {80}},\ \bibinfo {pages} {3923}
		(\bibinfo {year} {1996})}\BibitemShut {NoStop}%
	\bibitem [{\citenamefont {Nagel}\ \emph {et~al.}(2008)\citenamefont {Nagel},
		\citenamefont {Speer}, \citenamefont {Gaber}, \citenamefont {Sterck},
		\citenamefont {Eichhorn}, \citenamefont {Reimann}, \citenamefont {Ilin},
		\citenamefont {Siegel}, \citenamefont {Koelle},\ and\ \citenamefont
		{Kleiner}}]{NagelPRL}%
	\BibitemOpen
	\bibfield  {author} {\bibinfo {author} {\bibfnamefont {J.}~\bibnamefont
			{Nagel}}, \bibinfo {author} {\bibfnamefont {D.}~\bibnamefont {Speer}},
		\bibinfo {author} {\bibfnamefont {T.}~\bibnamefont {Gaber}}, \bibinfo
		{author} {\bibfnamefont {A.}~\bibnamefont {Sterck}}, \bibinfo {author}
		{\bibfnamefont {R.}~\bibnamefont {Eichhorn}}, \bibinfo {author}
		{\bibfnamefont {P.}~\bibnamefont {Reimann}}, \bibinfo {author} {\bibfnamefont
			{K.}~\bibnamefont {Ilin}}, \bibinfo {author} {\bibfnamefont {M.}~\bibnamefont
			{Siegel}}, \bibinfo {author} {\bibfnamefont {D.}~\bibnamefont {Koelle}},\
		and\ \bibinfo {author} {\bibfnamefont {R.}~\bibnamefont {Kleiner}},\
	}\bibfield  {title} {\bibinfo {title} {Observation of negative absolute
			resistance in a josephson junction},\ }\href
	{https://doi.org/10.1103/PhysRevLett.100.217001} {\bibfield  {journal}
		{\bibinfo  {journal} {Phys. Rev. Lett.}\ }\textbf {\bibinfo {volume} {100}},\
		\bibinfo {pages} {217001} (\bibinfo {year} {2008})}\BibitemShut {NoStop}%
	\bibitem [{\citenamefont {Gurvitch}\ \emph {et~al.}(1983)\citenamefont
		{Gurvitch}, \citenamefont {Washington},\ and\ \citenamefont
		{Huggins}}]{Gurvitch1983}%
	\BibitemOpen
	\bibfield  {author} {\bibinfo {author} {\bibfnamefont {M.}~\bibnamefont
			{Gurvitch}}, \bibinfo {author} {\bibfnamefont {M.~A.}\ \bibnamefont
			{Washington}},\ and\ \bibinfo {author} {\bibfnamefont {H.~A.}\ \bibnamefont
			{Huggins}},\ }\bibfield  {title} {\bibinfo {title} {High quality refractory
			josephson tunnel junctions utilizing thin aluminum layers},\ }\href
	{https://doi.org/10.1063/1.93974} {\bibfield  {journal} {\bibinfo  {journal}
			{Appl. Phys. Lett.}\ }\textbf {\bibinfo {volume} {42}},\ \bibinfo {pages}
		{472} (\bibinfo {year} {1983})}\BibitemShut {NoStop}%
	\bibitem [{\citenamefont {Lolli}\ \emph {et~al.}(2016)\citenamefont {Lolli},
		\citenamefont {Taralli}, \citenamefont {Portesi}, \citenamefont {Rajteri},\
		and\ \citenamefont {Monticone}}]{Lolli2016}%
	\BibitemOpen
	\bibfield  {author} {\bibinfo {author} {\bibfnamefont {L.}~\bibnamefont
			{Lolli}}, \bibinfo {author} {\bibfnamefont {E.}~\bibnamefont {Taralli}},
		\bibinfo {author} {\bibfnamefont {C.}~\bibnamefont {Portesi}}, \bibinfo
		{author} {\bibfnamefont {M.}~\bibnamefont {Rajteri}},\ and\ \bibinfo {author}
		{\bibfnamefont {E.}~\bibnamefont {Monticone}},\ }\bibfield  {title} {\bibinfo
		{title} {Aluminum–titanium bilayer for near-infrared transition edge
			sensors},\ }\href {https://doi.org/10.3390/s16070953} {\bibfield  {journal}
		{\bibinfo  {journal} {Sensors (Basel)}\ }\textbf {\bibinfo {volume} {16}},\
		\bibinfo {pages} {953} (\bibinfo {year} {2016})}\BibitemShut {NoStop}%
	\bibitem [{\citenamefont {Fornieri}\ \emph {et~al.}(2017)\citenamefont
		{Fornieri}, \citenamefont {Timossi}, \citenamefont {Virtanen}, \citenamefont
		{Solinas},\ and\ \citenamefont {Giazotto}}]{FornieriTimossi}%
	\BibitemOpen
	\bibfield  {author} {\bibinfo {author} {\bibfnamefont {A.}~\bibnamefont
			{Fornieri}}, \bibinfo {author} {\bibfnamefont {G.}~\bibnamefont {Timossi}},
		\bibinfo {author} {\bibfnamefont {P.}~\bibnamefont {Virtanen}}, \bibinfo
		{author} {\bibfnamefont {P.}~\bibnamefont {Solinas}},\ and\ \bibinfo {author}
		{\bibfnamefont {F.}~\bibnamefont {Giazotto}},\ }\bibfield  {title} {\bibinfo
		{title} {0–$\pi$ phase-controllable thermal josephson junction},\ }\href
	{https://doi.org/10.1038/nnano.2017.25} {\bibfield  {journal} {\bibinfo
			{journal} {Nature Nanotech.}\ }\textbf {\bibinfo {volume} {12}},\ \bibinfo
		{pages} {425} (\bibinfo {year} {2017})}\BibitemShut {NoStop}%
	\bibitem [{\citenamefont {De~Gennes}(1964)}]{DEGENNESRMP}%
	\BibitemOpen
	\bibfield  {author} {\bibinfo {author} {\bibfnamefont {P.~G.}\ \bibnamefont
			{De~Gennes}},\ }\bibfield  {title} {\bibinfo {title} {Boundary effects in
			superconductors},\ }\href {https://doi.org/10.1103/RevModPhys.36.225}
	{\bibfield  {journal} {\bibinfo  {journal} {Rev. Mod. Phys.}\ }\textbf
		{\bibinfo {volume} {36}},\ \bibinfo {pages} {225} (\bibinfo {year}
		{1964})}\BibitemShut {NoStop}%
	\bibitem [{Note1()}]{Note1}%
	\BibitemOpen
	\bibinfo {note} {The factor 2 takes into account the presence of the two
		junctions.}\BibitemShut {Stop}%
\end{thebibliography}
\end{document}